\documentclass[seceq,preprint]{ptptex}

\usepackage{bm}
\usepackage{cite}

\def\cL{\mathcal{L}}
\def\cM{\mathcal{M}}
\def\cO{\mathcal{O}}
\def\bR{\mathbb{R}}

\def\bL{\bm{L}}
\def\bTheta{\bm{\Theta}}
\def\bQ{\bm{Q}}
\def\bW{\bm{W}}
\def\bX{\bm{X}}
\def\bE{\bm{E}}
\def\bS{\bm{S}}
\def\bk{\bm{k}}
\def\bomega{\bm{\omega}}
\def\bOmega{\bm{\Omega}}
\def\bj{\bm{j}}
\def\cov{\text{cov}}
\def\interior{\mathop{\lrcorner}}
\def\lie{\pounds}
\def\vol{\mathrm{vol}}
\def\d{\partial}
\def\at{ {\itshape at} }
\def\email{email: }

\title{Higher-Derivative Corrections\\
 to the Asymptotic Virasoro Symmetry\\
of 4d Extremal Black Holes}
\author{
Tatsuo {\sc Azeyanagi}$^{\spadesuit,}$\footnote{\email aze\at gauge.scphys.kyoto-u.ac.jp},
Geoffrey {\sc Comp\`ere}$^{\heartsuit,}$\footnote{\email gcompere\at physics.ucsb.edu},\\
Noriaki {\sc Ogawa}$^{\diamondsuit,}$\footnote{\email noriaki\at yukawa.kyoto-u.ac.jp},
Yuji {\sc Tachikawa}$^{\clubsuit,}$\footnote{\email yujitach\at ias.edu},
and
Seiji {\sc Terashima}$^{\diamondsuit,}$\footnote{\email terasima\at yukawa.kyoto-u.ac.jp}
}

\inst{$^\spadesuit$
Department of Physics, Kyoto University,\\
Kyoto 606-8502, Japan, \\
$^\heartsuit$
Department of Physics, University of California at Santa Barbara,\\
Santa Barbara, CA 93106, USA,\\
$^\diamondsuit$
Yukawa Institute for Theoretical Physics, Kyoto University,\\
Kyoto 606-8502, Japan,\\
$^\clubsuit$
School of Natural Sciences, Institute for Advanced Study,\\
Princeton, NJ 08540, USA}

\abst{\large 
We study the asymptotic Virasoro symmetry
which acts on the near-horizon region of extremal four-dimensional black hole solutions of gravity theories 
with higher-derivative corrections, following the recently proposed Kerr/CFT 
correspondence.
We demonstrate that its central charge correctly reproduces 
the entropy formula of Iyer-Wald, once the boundary terms  in the symplectic structure
are carefully chosen.
}

\preprintnumber[3cm]{
\hbox{}\hfill arXiv:0903.4176\\
\hbox{}\hfill KUNS-2197\\
\hbox{}\hfill YITP-09-20}

\begin{document}
\maketitle
\section{Introduction}
One of the most surprising aspects of the black holes  is that the semi-classical analysis shows that they behave as thermodynamic systems\cite{Bekenstein:1973ur, Hawking:1974sw}. This begged a natural question how to account for their entropy in terms of statistical mechanics. 
One important breakthrough is the observation by Brown and Henneaux\cite{Brown:1986nw} that the asymptotic symmetries of the three-dimensional anti-de Sitter space (AdS$_3$) consist of two copies of the Virasoro algebra
with finite central charge, which is the symmetry group of the two-dimensional conformal field theories (CFTs). 
From the modern point of view, this was one of the earliest manifestations of the AdS/CFT correspondence\cite{Maldacena:1997re}, which relates the physics of the bulk AdS and the boundary CFT. Strominger then showed \cite{Strominger:1997eq} that the entropy of the three-dimensional black holes found by Ba\~nados-Teitelboim-Zanneli\cite{Banados:1992wn} can be obtained by applying the Cardy formula to this two-dimensional CFT.\footnote{Subsequently it was argued\cite{Carlip:1998wz,Solodukhin:1998tc} that the Virasoro symmetry can be found in any black hole horizons and that it reproduces the entropy correctly via the Cardy formula.}

Last year this approach was generalized to four-dimensional (4d) Kerr black holes by Guica, Hartman, Song and Strominger\cite{Guica:2008mu},
who found that the near-horizon geometry of the extremal Kerr black holes has one copy of a Virasoro algebra as its asymptotic symmetry under a judicious choice of the boundary condition on the fall-off of the metric. The central charge was calculated using  the formalism\cite{Lee:1990nz,Wald:1993nt,Iyer:1994ys,Anderson:1996sc,Barnich:2001jy,Barnich:2007bf,Compere:2007az}
which covariantizes the calculation done 
by Brown and Henneaux\cite{Brown:1986nw}.
Combined with the Frolov--Thorne temperature\cite{Frolov:1989jh} associated to the rotation,
the entropy of the extremal Kerr black holes was correctly reproduced. 
This observation ignited a flurry of activities to generalize the idea to other types of extremal black holes of various gravity theories with matter fields in various dimensions\cite{Hotta:2008xt,Lu:2008jk,Azeyanagi:2008kb,Hartman:2008pb,Nakayama:2008kg,Chow:2008dp,Isono:2008kx,Azeyanagi:2008dk,Peng:2009ty,Chen:2009xja,Loran:2009cr,Ghezelbash:2009gf,Lu:2009gj,Compere:2009dp,Astefanesei:2009sh,Garousi:2009zx}.
These works showed that what was crucial was the extremality and the $SL(2,\bR)\times U(1)$ symmetry of the background. Currently  we only know the central charge of the putative dual CFT;
firmer understanding of this dual CFT would lead us to what could be called the extremal black hole/CFT correspondence.

In these preceding works, 
the analyses were done for the Einstein-Hilbert action with and without matter fields.
However, when we regard
the gravity theory as the low-energy effective theory
of its ultraviolet completion such as string theory,
it is expected that the Lagrangian contains Planck-suppressed higher-derivative correction terms of the metric and other fields. 
They replace
the Bekenstein-Hawking entropy formula \begin{equation}
S= \frac1{4G_N \hbar }\int_\Sigma  \vol(\Sigma)
\end{equation}
by the Iyer-Wald entropy formula\cite{Wald:1993nt,Jacobson:1993vj,Iyer:1994ys} (the notation will be explained in detail in the next section)
\begin{equation}
S= -\frac{2\pi}{\hbar} \int_\Sigma \frac{\delta^\cov L}{\delta R_{abcd}} 
\epsilon_{ab} \epsilon_{cd}\vol(\Sigma). 
\end{equation} 
Our objective in this paper is then to show that this Iyer-Wald entropy for the extremal rotating black holes can be correctly reproduced by evaluating the central charge of the asymptotic Virasoro algebra in the presence of the higher-derivative corrections.\footnote{The higher-derivative contribution to the central charge of the asymptotic Virasoro algebra of AdS$_3$ was studied in \citen{Saida:1999ec,Hotta:2008yq}. 
The former treated the diffeomorphism-invariant Lagrangian density, but used the field redefinition specific to three dimensions which rewrites arbitrary such Lagrangians to the Einstein-Hilbert term with scalar fields with higher-derivative interactions. 
The latter paper dealt the topologically massive gravity\cite{DeserJackiwTempleton} in the canonical ADM formalism, more directly following the approach taken by Brown-Henneaux.\cite{Brown:1986nw}. It would be instructive to redo their analyses using the covariant phase space method.}  To simplify the calculation we introduce a tower of auxiliary fields so that the Lagrangian does not contain explicit derivatives higher than the second. 
We will see that the use of the symplectic structure and asymptotic charges advocated by
Barnich, Brandt and one of the authors\cite{Barnich:2001jy,Barnich:2007bf,Compere:2007az} is crucial in obtaining the agreement\footnote{The relationship between cohomological methods\cite{Anderson:1996sc,Barnich:2001jy,Barnich:2007bf,Compere:2007az} and the closely related covariant methods based on the linear equations of motion\cite{Abbott:1981ff,Deser:2002jk,Deser:2007vs} and covariant symplectic methods in first order theories\cite{Julia:1998ys,Julia:2002df} are detailed in \citen{Compere:2007az}.}.
For concreteness we work with arbitrary diffeomorphism-invariant four-dimensional Lagrangian whose only dynamical field is the metric, but the method we will employ is general enough to be applied
to any sensible Lagrangian. 
We expect that the analysis would also work for higher dimensional cases, by reducing the geometry to the four-dimensional one which we deal with in this paper; 
we expect that the Kaluza-Klein fields would not contribute to the central charge,  since the $U(1)$ gauge fields and scalar fields was shown not to contribute in the case of Einstein gravity \cite{Compere:2009dp}.

The structure of our paper is as follows: we start by recalling how the extremal black hole/CFT correspondence works in the absence of  higher-derivative corrections in Sec.~\ref{review}. We then review in Sec.~\ref{formalism} the method to determine the form of the asymptotic charges starting from the Lagrangian.  We then apply it in Sec.~\ref{charges} to the Lagrangian with higher-derivative corrections constructed from the metric. The resulting asymptotic charges will be evaluated on the extremal black hole background in Sec.~\ref{evaluation}, and we will see that the central charge perfectly reproduces the Iyer-Wald entropy. We will conclude with a short discussion in Sec.~\ref{summary}. 

There are a few appendices: App.~\ref{integrability} checks the integrability of the asymptotic charges for the Lagrangian with the Gauss-Bonnet term  and the finiteness of the Virasoro charges for a generic Lagrangian.
In App.~\ref{FT} we argue that the Frolov-Thorne temperature is not corrected in the presence of the higher-derivative terms. App.~\ref{homotopy} collects the formulae we use in the variational calculus. App.~\ref{tricks} details the constraint imposed on the tensors by the isometry of the near-horizon region of the extremal black hole. 

\section{Review of the extremal black hole/CFT correspondence}\label{review}
Let us start by reviewing how the calculation of the entropy
of extremal black holes works in terms of the asymptotic Virasoro symmetry\cite{Guica:2008mu}.
We will point out during the way which part needs to be modified in the presence of the higher-derivative terms in the Lagrangian. The overall presentation in this section will follow largely the one given in \citen{Chow:2008dp,Compere:2009dp}.

The extremal black hole is defined as the one  whose  inner and outer horizons coincide.
It implies the existence of the scaling symmetry in the near horizon region,
which is always automatically enhanced to the $SL(2,\bR)$ symmetry as shown in \citen{Kunduri:2007vf}.
We can choose a coordinate system
such that the near-horizon metric is given by 
\begin{equation}
ds^2=A(\theta)^2 \left(-r^2dt^2 + \frac{dr^2}{r^2}\right) + 
d\theta^2+ B(\theta)^2 (d\varphi + k r dt)^2.\label{metric}
\end{equation}   
Here $\varphi$ is an angular variable 
which takes values in
$0\le \varphi < 2\pi$.
The constant $k$ and the functions 
$A(\theta)$, $B(\theta)$ are determined by solving the equations of motion,
or using the entropy function formalism\cite{Astefanesei:2006dd,Astefanesei:2007bf}. 
As shown in \citen{Kunduri:2007vf}, this form is valid even in the presence
of higher-derivative corrections in the Lagrangian provided that the black hole is
big, in the technical sense that the curvature at the horizon remains finite
in the limit where the higher-derivative corrections vanish. 

This metric has the symmetry $SL(2,\bR)\times U(1)$ generated by 
\begin{align}
\zeta_1&=\partial_t,&\zeta_2&=t\partial_t-r\partial_r,&
\zeta_3&=\bigg(\frac{1}{2r^2}+\frac{t^2}{2}\bigg)\partial_t-tr\partial_r -\frac kr\partial_\varphi,  
& \zeta_0&=\partial_\varphi.
\label{killing}
\end{align}
It is also invariant under the discrete symmetry which maps\begin{equation}
(t,\varphi) \to (-t,-\varphi).
\end{equation} 
This is often called the $t$-$\varphi$ reflection symmetry in the black hole literature.

Following the argument in \citen{Guica:2008mu} we impose the boundary condition
\begin{equation}
\delta g_{\mu\nu} \sim 
\left(\begin{array}{cccr@{\,}l}
\delta g_{tt} = \cO(r^2) & \delta g_{tr}=\cO(r^{-2}) & \delta g_{t\theta}=\cO(r^{-1}) & \delta g_{t\varphi} &= \cO(1) \\
& \delta g_{rr} = \cO(r^{-3}) & \delta g_{r\theta}=\cO(r^{-2}) & \delta g_{r\varphi}&=\cO(r^{-1}) \\
&& \delta g_{\theta\theta}=\cO(r^{-1}) & \delta g_{\theta \varphi}&=\cO(r^{-1}) \\
&&& \delta g_{\varphi\varphi}&=\cO(1)
\end{array}\right)\label{bc}
\end{equation} on the metric, which is preserved by the vector fields \begin{equation}
\xi_n=-e^{-in\varphi }(\partial_\varphi + inr\partial_r), \label{Vf}
\end{equation} 
whose commutation relations are
\begin{equation}
i[\xi_m,\xi_n]=(m-n)\xi_{m+n}.
\label{xialgebra}
\end{equation} 
Here $m$ and $n$ are integers.

It is easy to check that they indeed preserve the boundary condition above, 
using the vierbein \begin{align}
e^{\hat t}&=A(\theta) r dt,&
e^{\hat r}&=A(\theta) dr/r,&
e^{\hat \theta}&= d\theta,&
e^{\hat \varphi}&= B(\theta)(d\varphi+krdt),\label{vierbein}
\end{align}  and their variation under $\xi_n$:
\begin{align}
\lie_{\xi_n} e^{\hat t}&= - e^{-in\varphi} i n e^{\hat t},&
\lie_{\xi_n} e^{\hat r} &= e^{in\varphi} n^2\Big(-k e^{\hat t}+ \frac AB  e^{\hat \varphi}\Big),\\
\lie_{\xi_n} e^{\hat \theta}&=0,&
\lie_{\xi_n} e^{\hat \varphi}&= in e^{-in\varphi} \Big(-\frac{2kB}{A} e^{\hat t}+e^{\hat \varphi}\Big). \label{derivatives-of-bases}
\end{align} Here $\lie_\xi$ denotes the Lie derivative by the vector field $\xi$. Components in the vierbein basis will be distinguished by hats on the indices in what follows.

We can always associate the charge $H_\xi$ to the asymptotic isometry $\xi$ at least formally. Whether it is well-defined depends on the boundary conditions. 
The charges  $H_n$, corresponding to the Virasoro symmetries $\xi_n$,
are finite in general as shown in Appendix \ref{integrability}. The boundary conditions (\ref{bc}) are also preserved by $\partial_{t}$. As a part of the boundary conditions, we impose the Dirac constraint $H_{\partial_t}=0$. We showed in Appendix \ref{integrability} that at least for Einstein 
gravity coupled to Gauss-Bonnet gravity the charges are integrable around the background but we do not have a proof of integrability around other solutions obeying the boundary conditions or for other  Lagrangians. 
We assume that integrability holds in what follows which can always be achieved using, if necessary, supplementary constraints. The charges $H_n$ then form a representation of the algebra \eqref{Vf} and they are conserved because their Dirac bracket with $H_{\partial_t}$ is zero and $\xi_n$ is time-independent.

The crucial observation\cite{Guica:2008mu} was that, just as in the case of $AdS_3$,\cite{Brown:1986nw} 
the Dirac bracket among the charges
$H_{n}$ acquires the central extension 
\begin{equation}
i \{ H_m,H_n \} =(m-n)H_{m+n} + \frac{c}{12} m(m^2+a) \delta_{m,-n},
\label{algebra}
\end{equation} which is the Virasoro algebra with central charge $c$.\footnote{
Here $a$ corresponds to a trivial cocycle and can be absorbed to a
redefinition of $H_0$. 
One can determine a natural definition of the angular momentum $H_0=H_{\partial_\varphi}$ 
by performing such change so that $a$ becomes the standard $-1$, but we do not pursue this direction
in this paper.}

When the Lagrangian is given purely by the Einstein-Hilbert term
\begin{equation}
\frac1{16\pi G_N}\int d^4x \sqrt{-g} R \label{EH},
\end{equation} the charges $H_{\zeta}$ is given by the formula
\begin{equation}
\delta H_\zeta =\int_{\Sigma} \bk_\zeta[\delta g; g],
\end{equation}
where $\Sigma$ is the sphere at the spatial infinity, and
\begin{multline}
\bk_\zeta[\delta g; g]=
\frac{1}{32\pi G_N}\epsilon_{a bcd }
\Bigl[
\zeta^{d }\nabla^{c }\delta g^{e }_{\ e } -\zeta^{d }\nabla_{e }\delta g^{c e }+
\zeta_{e }\nabla^{d }\delta g^{c e } \\
\quad +\frac{1}{2}\delta g^{e }_{\ e } \nabla^{d }\zeta^{c }
-\delta g^{d e }\nabla_{e }\zeta^{c }
+\frac{1}{2}\delta g^{e d }(\nabla^{c }\zeta_{e }+\nabla_{e }\zeta^{c })
\Bigr]dx^{a}\wedge dx^{b}.\label{kEH}
\end{multline}
Then the central term is given by 
\begin{equation}
\frac{c}{12}m(m^2+a) \delta_{m,-n} = i \int_\Sigma  \bk_{\xi_m}[\delta_{\xi_{n}}g;g],
\end{equation} and explicit evaluation shows 
\begin{equation}
c=\frac{3k}{2\pi G_N} \int_\Sigma d\theta d\varphi B(\theta) = \frac{3k}{2\pi G_N} \int_\Sigma \vol(\Sigma),
\end{equation}
where 
$\vol(\Sigma)=B(\theta)d\theta\wedge d\varphi$  is the natural volume form on the surface $\Sigma$. 
Using the correspondence principle mapping Dirac brackets $\{ .\,,.\}$ to commutators $\frac{i}{\hbar}[.\, , .]$, the dimensionless operators $L_m$ corresponding to $\frac{1}{\hbar} H_{m}$ then obey a Virasoro algebra with central charge 
\begin{equation}
\frac{c}{\hbar}= \frac{3k}{2\pi G_N \hbar } \int_\Sigma \vol(\Sigma). 
\end{equation}
A non-extremal black hole is in the ensemble weighted by the Boltzmann factor \begin{equation}
\exp\Big(-\frac{1}{T_H}(H-\Omega_H J)\Big).
\end{equation} In the extremal limit, it becomes
\begin{equation}
\exp\Big(-\frac1{T_\text{FT}}J\Big),
\end{equation}
where the Frolov-Thorne temperature $T_\text{FT}$ is given by
\begin{equation}
T_\text{FT}=\frac{1}{2\pi k}, \label{FTformula}
\end{equation}  where $k$ is the constant appearing in the metric \eqref{metric}. Notice that unlike the Bekenstein-Hawking temperature, no factor of $\hbar$ appears in this temperature.

Noticing that $J$ is the $H_0$ in the Virasoro algebra, one can apply the Cardy formula 
\begin{equation}
S=\frac{\pi^2}{3 \hbar } c\, T, \label{Cardy}
\end{equation}
where $T$ is the temperature of the CFT,
to obtain the entropy \begin{equation}
S= \frac1{4G_N \hbar }\int_\Sigma  \vol(\Sigma). \label{BHentropy}
\end{equation} 
Now we can move the surface $\Sigma$ to a finite value of $r$  without changing the integral, thanks to the scaling symmetry $\zeta_2$.  
Then $\Sigma$ can be identified with the horizon cross section of the extremal black hole.

Remarkably this expression \eqref{BHentropy} exactly reproduces the Bekenstein-Hawking entropy including the coefficient, which states that the entropy is proportional to the area of the horizon.
This original observation on the four-dimensional extremal Kerr black hole was soon extended to other extremal black holes for various theories in various dimensions.

If we think of the Lagrangian of the gravity theory as that of the low-energy effective theory of
string or M-theory which is a consistent ultraviolet completion of gravity,
it is expected that the Einstein-Hilbert Lagrangian \eqref{EH}
will have many types of Planck-suppressed higher-derivative corrections, and the total
Lagrangian is given by 
\begin{equation}
\int d^4x \sqrt{-g} f(g_{ab},R_{abcd},\nabla_e R_{abcd},\cdots ),
\end{equation} where $f$ is a complicated function.
The higher-derivative terms correct the black hole entropy in two ways: 
one by modifying the solution 
through the change in the equations of motion,
the other by  correcting the Bekenstein-Hawking area formula \eqref{BHentropy}
to the Iyer-Wald entropy formula 
\begin{equation}
S= -\frac{2\pi}{\hbar } \int_\Sigma \frac{\delta^\cov f}{\delta R_{abcd}} 
\epsilon_{ab} \epsilon_{cd}\vol(\Sigma). \label{Wald}
\end{equation} Here $\Sigma$ is the horizon cross section, and $\epsilon_{ab}$ is the binormal to the horizon, i.e.~the standard volume element of the normal bundle to $\Sigma$.
${\delta^\cov}/{\delta R_{abcd}}$ is the covariant Euler-Lagrange derivative of the Riemann tensor defined as
\begin{equation}
\frac{\delta^\cov}{\delta R_{abcd}} =  \sum_{i= 0} (-1)^i \nabla_{(e_1}\dots \nabla_{e_i)}\frac{\d  }{\d \nabla_{(e_1}\dots \nabla_{e_i)}R_{abcd}}. \label{defD}
\end{equation}
Naively, this is obtained by varying the Lagrangian with respect to
the Riemann tensor as if it were an independent field.

Our aim in this paper is to show that the Iyer-Wald formula is reproduced from the
consideration of the central charge of the boundary Virasoro algebra.
In order to carry it out, we first need to know how the asymptotic charge \eqref{kEH}
gets modified by the higher-derivative corrections.
Therefore we  now have to reacquaint ourselves how the asymptotic charges
and the central charge in their commutation relations are determined  for a given Lagrangian.

\section{Formalism}\label{formalism}
\subsection{The covariant phase space}
Let us begin by recalling how to construct the covariant phase space \cite{Lee:1990nz}.
We denote the spacetime dimension by $n$.
The input is the Lagrangian $n$-form $\bL=\star L$ which is a local functional
of fields $\phi^i$. Here $\phi^i$ stands for all the fields, including the metric.
$\star$ is the Hodge star operation, and $L$ is the Lagrangian density in the usual sense.
The equation of motion $(\text{EOM})_i$ for the field $\phi^i$ is determined
by taking the variation of $\bL$ and using the partial integration:
\begin{equation}
\delta \bL = (\mathrm{EOM})_i \delta\phi^i + d\bTheta. \label{eq:varL}
\end{equation} Here and in the following, 
we think of $\delta\phi^i$ as a one-form on the space of field configuration, 
just as the proper mathematical way to think of $dx^\mu$ is 
not just as an infinitesimal displacement but as a one-form on the spacetime.

The equation above does not fix 
the ambiguity of $\bTheta$ of the form $\bTheta\to \bTheta + d\bm Y$.
We fix it by defining $\bTheta$ by $\bTheta=- I_{\delta \phi}^n \bL$,
where the homotopy operator $I^n_{\delta\phi}$ is 
defined in Appendix~\ref{homotopy}.\footnote{
The definition for $\bTheta$ is precisely the minus the definition given in (2.12) of Lee-Wald\cite{Lee:1990nz}. Our minus sign comes from the convention $\{d,\delta\} = 0$, see Appendix~\ref{homotopy}.} 
The symplectic structure of the configuration space, 
as defined in Lee-Wald\cite{Lee:1990nz},  is then given by the integral of 
\begin{equation}
\bomega^{LW} =\delta \bTheta \label{omegaIW}
\end{equation} 
over the Cauchy surface $C$, 
\begin{equation}
\bOmega^{LW}[\delta_1\phi,\delta_2\phi; \phi]=\int_C \bomega^{LW}[\delta_1\phi,\delta_2\phi; \phi].
\end{equation}

One particularity of this construction is the non-invariance under the change of the Lagrangian by a total derivative term $\bL\to \bL+d\cL$ which does not change the equations of motion.
This induces the change
\begin{equation}
\bomega^{LW} \to \bomega^{LW}+ d\bomega_{\cL},
\end{equation} 
where $\bomega_{\cL} = \delta I^{n-1}_{\delta \phi}{\cL}$ is determined by the boundary term $\cL$. When the spatial directions are closed, or the asymptotic
fall-off of the fields is sufficiently fast, this boundary term does not contribute to the symplectic structure, but we need to be more careful in our situation where the boundary conditions \eqref{bc} allow 
$\cO(1)$ change with respect to the leading term. It was advocated in \citen{Barnich:2007bf,Compere:2007az} based on the cohomological results of \citen{Barnich:2001jy} to replace the definition \eqref{omegaIW} by the so-called invariant symplectic structure\footnote{This definition corresponds to the one advocated in \citen{Julia:2002df} in first order theories. In general, boundary terms should be added to the action to make it a well-defined variational principle. As argued in \citen{Compere:2008us}, if these boundary terms contain derivatives of the fields, they will contribute in general to a boundary term in the symplectic structure. We will not look at these additional contributions here. Our result indicates that these boundary terms, if any, do not contribute to the Virasoro central charge.}
\begin{equation}
\bomega^{inv} = -\frac{1}{2} I_{\delta \phi}^n \left( \delta \phi^i \frac{\delta \bL}{\delta \phi^i} \right),
\end{equation}
which depends only on the equations of motion of the Lagrangian. This symplectic structure differs from the Lee-Wald symplectic structure \eqref{omegaIW} by a specific boundary term $\bE$ 
\begin{equation}
\bomega^{inv}=\bomega^{W} - d\bE,
\end{equation}
where $\bE$ is given by
\begin{equation}
\bE =-\frac12 I^{n-1}_{\delta\phi} \bTheta. 
\end{equation}

\subsection{The Noether charge}
Now suppose the Lagrangian is diffeomorphism invariant: 
\begin{equation}
\delta_\xi \bL=\lie_\xi \bL = d(\xi\interior \bL),
\end{equation} 
where $\xi$ is a vector field which generates an infinitesimal diffeomorphism
and $\lie_\xi$ is the Lie derivative with respect to $\xi$.
The corresponding Noether current is
\begin{equation}
\bj_\xi = -\bTheta[\lie_\xi\phi ; \phi] - \xi \interior \bL.
\end{equation} 
Here $\xi\interior$ stands for the interior product of a vector to a differential form, and 
\begin{equation}
\bTheta[\lie_\xi\phi ; \phi]\equiv \Big(\lie_\xi \phi \frac{\partial}{\partial\phi}+ \partial_a \lie_\xi \phi \frac{\partial}{\partial \phi_{,a}} + \cdots \Big) \interior \bTheta,
\end{equation}
that is, $\phi\to \phi+ \epsilon\lie_\xi\phi$ defines a vector field on the configuration space of $\phi$ and its derivatives $\phi_{,a}$, \dots, and we contract this vector field to the one-forms  $\delta\phi$, $\delta \phi_{,a}$  inside $\bTheta$, see Appendix \ref{homotopy} for more details. 

Using the Noether identities, one can write
\begin{equation}
d \bj_\xi = - \frac{\delta \bL}{\delta \phi^i}\lie_\xi \phi^i = d \bS_\xi,
\end{equation}
where $\bS_\xi$ is the on-shell vanishing Noether current. Since $\bj_\xi - \bS_\xi$ is off-shell closed and thus exact, there is a $(n-2)$-form $\bQ_\xi$ such that
\begin{equation}
\bj_\xi = d \bQ_\xi
\end{equation} on shell.  
This object $\bQ_\xi$ is the Noether charge as defined by Wald,\cite{Wald:1993nt} 
which when integrated over the bifurcate horizon gives the Iyer-Wald entropy.
This is closely related to the charge $H_\xi$
which generates the action of the diffeomorphism 
$\xi$ on the covariant phase space. 
By definition, the Hamiltonian which generates the flow $\phi^i \to \phi^i + \epsilon \delta_\xi \phi^i$ needs to satisfy 
\begin{equation}
\bOmega[\delta_\xi \phi, \delta\phi; \phi] = \delta H_\xi.
\end{equation}
Now let us define 
\begin{equation}
\bk^{IW}_\xi[\delta\phi ; \phi] = \delta \bQ_\xi -\xi \interior \bTheta. \label{kIW}
\end{equation} 
Then one can show 
\begin{equation}
\bomega^{IW}[\delta_\xi\phi, \delta \phi; \phi] = d\bk^{IW}_\xi[\delta\phi ; \phi],
\end{equation} 
when $\phi$ solves the equations of motion and $\delta\phi$ solves the linearized equations of motion around $\phi$. 
Integrating over the Cauchy surface, we have 
\begin{equation}
\bOmega^{IW}[\delta_\xi\phi,\delta\phi; \phi]= \int_{\Sigma} \bk^{IW}_\xi[\delta\phi ; \phi],
\end{equation} 
where $\partial C = \Sigma$. Therefore we have 
\begin{equation}
\delta H_\xi^{IW} = \int_{\Sigma}\bk^{IW}_\xi,
\end{equation} when such $H_\xi$ exists.
The Hamiltonian is defined as
\begin{equation}
H^{IW}_\xi = \int_{\bar \phi}^{\phi}  \int_{\Sigma} \bk^{IW}_\xi[\delta\phi ; \phi].
\end{equation}
where the first integration is performed in configuration space between the reference solution $\bar \phi$ and $\phi$.   For this definition to be independent of the path in the configuration space,
the integrability conditions \begin{equation}
\int_\Sigma \delta \bk^{IW}_\xi=0
\end{equation} need to be obeyed, see Appendix \ref{integrability} for an analysis in Gauss-Bonnet gravity.

When one chooses $\bomega^{inv}$ instead as the symplectic form, 
one finds \begin{equation}
\bomega^{inv}[\delta_\xi\phi, \delta\phi; \phi] = d\bk^{inv}_\xi[\delta\phi ; \phi],
\end{equation} where 
\begin{equation}
\bk^{inv}_\xi[\delta\phi ; \phi] = \bk^{IW}_\xi[\delta\phi; \phi] - \bE[\delta_\xi\phi,\delta\phi; \phi]. \label{kBC}
\end{equation}
We can finally write down the formula for the representation of the asymptotic symmetry algebra by a Dirac bracket \cite{Barnich:2001jy,Koga:2001vq,Barnich:2007bf}: 
\begin{equation}
\delta_{\xi} H_\zeta\equiv  \{H_\zeta, H_\xi\}=H_{[\zeta,\xi]} +\int_{\Sigma} \bk_\zeta[\delta_\xi\phi ; \bar \phi]
\label{PB},
\end{equation}
which is valid on-shell when the conditions of integrability of the charges as well as the cocycle condition $\int_\Sigma \delta\bE = 0$ are obeyed.

Therefore, our task is to obtain the formula for $\bk_\xi^{IW,inv}$  for a general class of theories, and to evaluate the central charge given by \eqref{PB}. Before proceeding, let us recall that the form \eqref{kEH} for the asymptotic charge
of the Einstein-Hilbert theory corresponds to $\bk^{inv}$; the last term in \eqref{kEH} comes from $\bE$\footnote{  In the case of Einstein gravity, one can show using the linearized constraint equations described in the Appendix A of \citen{Guica:2008mu} that the components of $\bE[\delta_1 g,\delta_2 g;\bar g]$ tangent to $\Sigma$ vanish at the boundary $r \rightarrow \infty$ around the background $\bar g$ when we take it to be the near horizon of the extremal Kerr black hole. 
There is therefore no distinction between the on-shell invariant symplectic structure/charges and the Iyer-Wald symplectic structure/charges for Einstein gravity around $\bar g$.}.

\subsection{The central term}
Iyer and Wald\cite{Iyer:1994ys} showed that $\bQ_\xi$ has the form\footnote{ The ambiguities in $\bQ_\xi$ described in \citen{Iyer:1994ys} can be entirely fixed by defining $\bQ_\xi = I_\xi^{n-1} \bj_\xi = - I^{n-1}_\xi \bTheta[\lie_\xi \phi ; \phi]$, see Appendix \ref{homotopy} for definitions. The Noether charge for a general diffeomorphism invariant theory of gravity derived in Sec.~\ref{charges} will then precisely have this form.}
\begin{equation}
\bQ_\xi = \bW_c \xi^c +  \bX_{cd}\nabla^{[c} \xi^{d]}, \label{Qgeneral}
\end{equation} where $\bW_c$ and $\bX_{cd}$ are $(n-2)$-forms 
with extra indices $c$ and $(c,d)$ respectively, both  
covariant tensors constructed from $\phi$.
Moreover \begin{equation}
(\bX_{cd})_{c_3\cdots c_n}=-\epsilon_{abc_3\cdots c_n} Z^{ab}{}_{cd}, \label{XZ}
\end{equation}where $Z^{abcd}$ is defined by the relation \begin{equation}
\delta \bL =\star  Z^{abcd} \delta R_{abcd} +\cdots,
\end{equation} which is obtained by taking the functional derivative of $L$ 
with respect to the Riemann tensor $R_{abcd}$ as if it were an independent field: \begin{equation}
Z^{abcd}=\frac{\delta^{\cov} L}{\delta R_{abcd}}.
\end{equation}

Let us now massage the central term into a more tractable form:
\begin{align}
\int_\Sigma \bk^{IW}_\zeta[\lie_\xi\phi ; \bar\phi] 
&= \int_\Sigma \left[\delta_\xi \bQ_\zeta + \zeta\interior \bTheta(\lie_\xi \phi ; \bar\phi) \right]\\
&= \int_\Sigma \left[\delta_\xi \bQ_\zeta - \zeta\interior( d\bQ_\xi + \xi\interior \bL) \right] \\
&= \int_\Sigma \left[(\delta_\xi -\lie_\xi) \bQ_\zeta  +(\lie_\xi \bQ_\zeta - \lie_\zeta \bQ_\xi ) - \zeta\interior  \xi\interior\bL \right].
\end{align} In the last equality we used the fact $\lie_\zeta = d\zeta\interior + \zeta\interior d$.
Now the antisymmetry in $\zeta$ and $\xi$ 
is manifest except the first term in the last line. So let us deal with it.

We have the relations \begin{align}
\delta_\xi \bQ_\zeta &= \lie_\xi (\bW_c) \zeta^c + \lie_\xi (\bX_{cd}) \nabla^{[c} \zeta^{d]} 
+ \bX_{cd} \delta_\xi( \nabla^{[c} \zeta^{d]} ) ,\\
\lie_\xi \bQ_\zeta & =   \lie_\xi (\bW_c) \zeta^c + \bW_c [\xi,\zeta]^c + 
\lie_\xi (\bX_{cd}) \nabla^{[c} \zeta^{d]} 
+ \bX_{cd} \lie_\xi( \nabla^{[c} \zeta^{d]} ).
\end{align} We also know\footnote{This equation means that 
the covariant derivative of a vector $\zeta$
transforms as a tensor under the diffeomorphism generated by $\xi$, 
if the metric and the vector are both transformed 
by the diffeomorphism generated by $\xi$. 
The first and the second term on the right hand side are the changes induced
by the metric and by the vector, respectively.} 
\begin{equation}
\lie_\xi(\nabla^{[c}\zeta^{d]}) = \delta_\xi (\nabla^{[c}\zeta^{d]}) + \nabla^{[c}[\xi,\zeta]^{d]}.
\end{equation} 
Thus we have \begin{equation}
(\delta_\xi-\lie_\xi)\bQ_\zeta= 
-\bW_c[\xi,\zeta]^c - \bX_{cd} \nabla^{[c}[\xi,\zeta]^{d]} = \bQ_{[\zeta,\xi]},
\end{equation} so that \begin{equation}
\int_\Sigma\bk^{IW}_\zeta[\lie_\xi\phi;\bar\phi] 
= \\
\int_\Sigma \left[
\bQ_{[\zeta,\xi]}
-(\lie_\zeta \bQ_\xi -\lie_\xi \bQ_\zeta) -\zeta\interior \xi\interior \bL
\right].
\end{equation}  
Now the antisymmetry in $\xi$ and $\zeta$ is manifest. Using \eqref{kBC},
one finds \begin{equation}
\int_\Sigma \bk^{inv}_\zeta[\lie_\xi\phi;\bar\phi] 
= \\
\int_\Sigma \left[
\bQ_{[\zeta,\xi]}
-(\lie_\zeta \bQ_\xi -\lie_\xi \bQ_\zeta) -\zeta\interior \xi\interior \bL
-\bE[\delta_\zeta\phi,\delta_\xi\phi ; \bar \phi]
\right].\label{kxi}
\end{equation}
 The first  term
on the right-hand side is a trivial cocycle since it can be absorbed into a shift of the Hamiltonian $H_{\zeta,\xi}$ in \eqref{PB}.

\section{Explicit form of charges for higher-derivative Lagrangian}\label{charges}
Our aim is to evaluate the central term reviewed in the last section 
on the extremal black hole background. We first need to have an explicit form of $\bTheta$,
$\bQ_\xi$ and $\bE$  for the higher-derivative Lagrangian, which we will carry out in this section.

\subsection{Lagrangians without derivatives of Riemann tensor}\label{without}
Let us first consider a Lagrangian of the form 
\begin{equation}
\bL=\star f(g_{ab}, R_{abcd}),
\label{L1}
\end{equation} 
where $f$ does not contain explicit derivatives. 
One can rewrite it as \begin{equation}
\bL=\star \left[f(g_{ab},\bR_{abcd}) + 
Z^{abcd} ( R_{abcd} - \bR_{abcd}) \right],
\label{higher}
\end{equation} where $\bR_{abcd}$ and $Z^{abcd}$
are auxiliary fields. Indeed, the variation of $\bR_{abcd}$ gives
\begin{equation}
Z^{abcd} = \frac{\partial f(g_{ab},\bR_{abcd})}{\partial \bR_{abcd}}, \label{onshellZnoD}
\end{equation} on-shell, while the variation of $Z^{abcd}$ gives \begin{equation}
R_{abcd}=\bR_{abcd}.
\end{equation} Therefore the Lagrangians \eqref{L1} and \eqref{higher} are equivalent.

Now that the Lagrangian does not have derivatives higher than the second derivative
of $g_{ab}$ contained in the Riemann tensor, so the calculation of 
$\bTheta$ etc.~is quite straightforward, and we have
\begin{equation}
\bTheta_{a_2\cdots a_n}= -2(Z^{abcd} \nabla_d \delta g_{bc} 
-(\nabla_d Z^{abcd})\delta g_{bc} )\epsilon_{aa_2\cdots a_n},
\label{Theta}
\end{equation} and \begin{equation}
(\bQ_\xi)_{c_3c_4\cdots c_n}=
 (-Z^{abcd}\nabla_c \xi_d - 2 \xi_c \nabla_d Z^{abcd}) \epsilon_{abc_3c_4\cdots c_n}.
\end{equation}
Comparing with \eqref{Qgeneral}, we see that 
\begin{equation}
(\bW^c)_{c_3\cdots c_{n}}=-2 \nabla_d Z^{abcd} \epsilon_{abc_3\cdots c_n} = 2 (\nabla_d \bX^{cd})_{c_3\cdots c_{n}}. \label{defW}
\end{equation}
The $\bE$ is obtained from the homotopy as argued above, and is
\begin{equation}
\bE_{a_3\cdots a_n}= \frac{1}{2} (-\frac{3}{2} Z^{abcd}\delta g_{c}{}^{ e}\wedge \delta g_{ed} +2  
Z^{acde} \delta g_{cd} \wedge \delta g^{b}{}_{e} )\epsilon_{ab a_3\cdots a_n}.\label{Eterm}
\end{equation}
Here we notice that
there is no term involving $\delta Z$.

\subsection{Lagrangians with  derivatives of Riemann tensor}
Generalization to Lagrangians with derivatives of Riemann tensor is also straightforward.
Take the Lagrangian \begin{equation}
\bL=\star f(g_{ab}, R_{abcd},\nabla_{e_1}R_{abcd},\nabla_{(e_1} \nabla_{e_2 )} R_{abcd}, \dots,\nabla_{(e_1}\dots \nabla_{e_k )} R_{abcd}),\label{higherD0}
\end{equation}
depending up to  $k$-th derivatives of the Riemann tensor.
This is the most general diffeomorphism-invariant Lagrangian density constructed from the metric
as was shown in \citen{Iyer:1994ys}. For example, any antisymmetric part of the covariant derivatives can be rewritten using the Riemann tensor with fewer number of derivatives.
As noted by  Iyer-Wald\cite{Iyer:1994ys} and Anderson-Torre\cite{Anderson:1996sc},
the tensors $\nabla_{(e_1}\cdots \nabla_{e_s)} R_{abcd}$ cannot be specified independently at a point because of differential identities satisfied by the curvature. The form of the Lagrangian is therefore not unique and has to be further specified. We assume in what follows that such a choice has been made. 

Now, one can introduce auxiliary fields and rewrite it as
\begin{eqnarray}
\bL&=&\star [f(g_{ab},\bR_{abcd}, \bR_{abcd | e_1},\dots,\bR_{abcd | e_1\dots e_k}) +  Z^{abcd} ( R_{abcd} - \bR_{abcd}) \nonumber \\
&&+  Z^{abcd|e_1} ( \nabla_{e_1} \bR_{abcd} - \bR_{abcd| e_1}) + 
Z^{abcd|e_1e_2} ( \nabla_{(e_2} \bR_{abcd|e_1)} - \bR_{abcd|e_1 e_2}) \nonumber\\
&&  +\dots + Z^{abcd|e_1\dots e_k} ( \nabla_{(e_k} \bR_{abcd|e_1\dots e_{k-1})} - \bR_{abcd|e_1 \dots e_{k}}) ].
\label{higherD}
\end{eqnarray} 
Here, the auxiliary fields $\bR_{abcd|e_1\dots e_s}$ and $Z^{abcd|e_1\dots e_s}$ for $ 1 \leq s \leq k$ are totally symmetric in the indices $e_1\dots e_s$ and the symmetrization in terms of the form $\nabla_{(e_s} \bR_{abcd|e_1\dots e_{s-1})}$ is among the $e_i$ indices only. Notice that $f$ does not contain explicit derivatives of the fields and that the only term with two derivatives is the one containing the Riemann tensor. 

The equations of motion for $\bR_{abcd|e_1\dots e_s}$ and $Z^{abcd|e_1\dots e_s}$ read as 
\begin{align}
\bR_{abcd|e_1\dots e_s} &= \nabla_{(e_s} \bR_{abcd|e_1\dots e_{s-1})}, \\
Z^{abcd|e_1\dots e_s} &=  \frac{\partial f}{\partial \bR_{abcd|e_1\dots e_s}} - \nabla_{e_{s+1}} Z^{abcd|e_1\dots e_{s+1}},
\end{align}
where for $s=0$ and $s=k$, there is no derivative term in the right-hand side of the second expression. These equations can be solved iteratively. One obtains in particular, \begin{eqnarray}
\bR_{abcd|e_1\dots e_s} &=& \nabla_{(e_1}\cdots \nabla_{e_s)} R_{abcd}, \\
Z^{abcd} &= & \frac{\delta^{\cov} }{\delta R_{abcd}}f(g_{ab}, R_{abcd},\nabla_{e_1}R_{abcd},\cdots), \label{onshellZ}
\end{eqnarray}
where the covariant Euler-Lagrange derivative of the Riemann tensor was defined in \eqref{defD}. Therefore, the Lagrangian \eqref{higherD} is equivalent to \eqref{higherD0}. 

The conserved charges for the Lagrangian \eqref{higherD} are simply the sum of the conserved charges for the Lagrangian \eqref{higher} with the on-shell condition \eqref{onshellZ} in place of \eqref{onshellZnoD} plus the conserved charges for the new terms with  $1 \leq s \leq k$ given by
\begin{eqnarray}
\bL^{(s)}&=& Z^{abcd|e_1\dots e_s} ( \nabla_{(e_s} \bR_{abcd|e_1\dots e_{s-1})} - \bR_{abcd|e_1 \dots e_{s}}).
\label{higherDs}
\end{eqnarray} 
Since the Lagrangian $\bL^{(s)}$ is only of first order in the derivatives of the fields, the correction terms to $\bTheta$ will contain no derivative. 
The full term $\bTheta$ is therefore given in \eqref{Theta} where $Z^{abcd}$ is \eqref{onshellZ} plus $k$ terms $\bTheta^{(s)}[\delta\phi; \phi]$ that we will compute soon. 
Since the $\bE$ term is obtained by a contracting homotopy $I_{\delta \phi}^{n-1}$ acting on the derivatives of the fields in $\bTheta$, there is no contribution to $\bE$ and \eqref{Eterm} is the final expression. Finally, the Noether charge $\bQ_{\xi} $ will contain only correction terms proportional to $\xi$, so we have contributions only to $\bW_c$ \eqref{defW}. Thus we conclude that $\bX_{cd}$ is indeed given by \eqref{XZ} as proven in \citen{Iyer:1994ys}. 

The outcome of this discussion is that we only have to compute the correction terms $\bTheta^{(s)}[\delta\phi; \phi]$, $\bW^{(s)}_c$ for each $1 \leq s \leq k$ coming from the Lagrangian \eqref{higherDs}. Application of the  homotopy operators then yields the results
\begin{eqnarray}
\bTheta^{(s)}_{a_2\cdots a_n}&= & \big( 2 ( Z^{ibcd|e_1 \dots e_{s-1} a} + Z^{abcd|e_1 \dots e_{s-1} i})\delta g_{ij} \bR^j_{\;\, bcd|e_1 \dots e_{s-1}}
 - 2 Z^{ibcd|e_1 \dots e_{s-1} j}\delta g_{ij} \bR^a_{\;\, bcd|e_1 \dots e_{s-1}} \nonumber\\
&& + (s-1) ( Z^{kbcd|e_1 \dots e_{s-2} ia} \delta g_{ij} \bR^{\quad\;\;\; j}_{kbcd|\;\,e_1 \dots e_{s-2}}
-\frac{1}{2} Z^{kbcd|e_1 \dots e_{s-2}i j} \delta g_{ij} \bR^{\quad\;\;\; a}_{kbcd|\;\,e_1 \dots e_{s-2}}) \nonumber\\
&& - Z^{kbcd|e_1 \dots e_{s-1} a}\delta \bR_{k bcd|e_1 \dots e_{s-1}} 
\big) \epsilon_{aa_2\cdots a_n},\label{Thetas}
\end{eqnarray} 
and 
\begin{multline}
(\bQ^{(s)}_\xi)_{c_3c_4\cdots c_n}=
-2 \xi_k \big(    Z^{klcd|e_1 \dots e_{s-1}a} \bR^{b}_{\;\,lcd|e_1 \dots e_{s-1}} + 
Z^{alcd|e_1 \dots e_{s-1}b}  \bR^{k}_{\;\,lcd|e_1 \dots e_{s-1}} \\
+ Z^{alcd|e_1 \dots e_{s-1}k}  \bR^{b}_{\;\,lcd|e_1 \dots e_{s-1}}
+\frac{s-1}{2} Z^{lmcd|e_1 \dots e_{s-2}ka}  \bR^{\quad\;\;\; b}_{lmcd|\;\,e_1 \dots e_{s-2}} \big)  \epsilon_{abc_3c_4\cdots c_n}. \label{Qs}
\end{multline}

\section{Central charge of the asymptotic Virasoro algebra}\label{evaluation}
Now we are finally in the position to evaluate 
the central extension for the algebra \eqref{xialgebra}
of the vector fields \eqref{Vf} on the background \eqref{metric}.
The central term \eqref{kxi} is easily shown to be a cocycle. Indeed, since the expression is manifestly anti-symmetric, it contains only odd powers of $n$. Moreover, because each Lie derivative can only generate two powers of $n$, the expression is at most quartic in $n$. There can therefore only be terms proportional to  $n$ and $n^3$.

To determine the central charge, it is sufficient to obtain the term proportional to $n^3$ in it.
Since  $i[\xi_{n},\xi_{-n}]=2n \xi_0$ and ${\xi_n}\interior \xi_{-n}\interior =2 inr \partial_r\interior  \partial_\varphi \interior$ are both only proportional to $n$, we have \begin{align}
\int_\Sigma \bk^{IW}_{\xi_n}[\lie_{\xi_{-n}}\phi ; \bar\phi] \bigm|_{n^3}
&= -\int_\Sigma (\lie_{\xi_n} \bQ_{\xi_{-n}} -\lie_{\xi_{-n}} \bQ_{\xi_{n}}) \bigm|_{n^3}\\
&=-2 \int_\Sigma \lie_{\xi_n} \bQ_{\xi_{-n}} \bigm|_{n^3} \\
&=-2 \int \left.\left[ \bX_{cd} \lie_{\xi_n} \nabla^c \xi_{-n}^d +  (\lie_{\xi_n} \bX)_{cd} \nabla^{[c} \xi_{-n}^{d]}
+ \lie_{\xi_n}\bW_c \,\xi^{c}_{-n}  \right]\right|_{n^3}.
\label{formula}
\end{align} where $|_{n^3}$ stands for the operation of extracting the term of order $n^3$.
In the following the placement of the indices are very important.
Since the vectors $\xi_n$ is only asymptotically Killing and moreover
it gives $\cO(1)$ contribution,  the Lie derivative with respect to $\xi_n$ does not commute
with the lowering/raising of the indices.

Let us evaluate the three terms in  \eqref{formula} in turn.
For simplicity, we first deal  with the Lagrangian without the derivatives of the Riemann tensor
discussed in Sec.~\ref{without}. We come back to the generalization to the Lagrangian with the derivatives of the Riemann tensor later in Sec.~\ref{later}.

\subsection{The first term}
Explicit evaluation of $\lie_{\xi_n} \nabla^c \xi_{-n}^d$ 
shows that the only $\cO(n^3)$ contribution in the first term of \eqref{formula} is in the 
$[cd]=[\hat t \hat r]$ and $=[\hat r\hat \varphi]$ components.
The integral gives terms proportional to 
$X_{\hat t\hat r | \hat\theta\hat\varphi}\propto Z_{\hat t\hat r \hat t \hat r}$ 
and $X_{\hat r\hat\varphi | \hat \theta\hat\varphi}\propto Z_{\hat r\hat \varphi \hat t \hat r }$
respectively. 
Now, the tensor $Z_{\hat r\hat \varphi \hat t \hat r }$ 
is zero due to the invariance of the metric under $SL(2,\bR)\times U(1)$,
see Appendix~\ref{tricks} for the details.
Therefore one finds 
\begin{align}
-2\int_\Sigma \bX_{cd} \lie_{\xi_n} \nabla^c \xi_{-n}^d  \bigm|_{n^3}
= 4i n^3 k \int_\Sigma Z_{\hat t\hat r \hat t \hat r}  \vol(\Sigma) 
= i n^3 k \int_\Sigma Z_{abcd} \epsilon^{ab}\epsilon^{cd}\vol(\Sigma),
\end{align}  where $\vol(\Sigma)=B(\theta)d\theta d\varphi$.
The contribution of the first term to the central charge is then \begin{equation}
c_\text{1st term}= -12k \int_\Sigma  Z_{abcd} \epsilon^{ab}\epsilon^{cd}\vol(\Sigma).
\end{equation} 
We show that there is no correction to the formula 
for the Frolov-Thorne temperature \eqref{FTformula} in Appendix~\ref{FT}. Then the application of the
Cardy formula gives that the contribution to the entropy from the first term is 
\begin{equation}
S_\text{1st term}= \frac{\pi^2}{3\hbar } c_\text{1st term} T_{FT}
=-\frac{ 2 \pi}{\hbar} \int_\Sigma  Z_{abcd} \epsilon^{ab}\epsilon^{cd}\vol(\Sigma),
\end{equation} which is exactly the celebrated formula of Iyer-Wald, \eqref{Wald}.
Therefore, our remaining task is to show that the rest of the terms in the central charge cancel among themselves.

\subsection{The second term}
In the following we will find it 
convenient to perform the Lie derivative in the vierbein components:
Let us define $\zeta^{\hat i}_{,\hat j}$ for a vector $\zeta$ via
\begin{equation}
\lie_\zeta e^{\hat i}=\zeta^{\hat i}_{,\hat j} e^{\hat j}.
\end{equation} Then we have \begin{equation}
(\lie_\zeta T)_{\hat a\hat b\hat c\cdots} =
\zeta^s \partial_sT_{\hat a\hat b\hat c\cdots} +
\zeta^{\hat i}_{,\hat a} T_{\hat i \hat b \hat c\cdots}+
\zeta^{\hat i}_{,\hat b} T_{\hat a\hat i  \hat c\cdots}+\cdots.\label{lie-formula}
\end{equation} $(\xi_{n})^{\hat i}{}_{,\hat j}$ can be read off from \eqref{derivatives-of-bases}.

The second term of \eqref{formula} is 
\begin{equation}
-2\int_\Sigma (\lie_{\xi_n} \bX)_{cd} \nabla^{[c} \xi_{-n}^{d]}
= -\int_\Sigma (\lie_{\xi_n} \bX)_{cd | c_3c_4} \nabla^{[c} \xi_{-n}^{d]} dx^{c_3}dx^{c_4}.
\end{equation}
Now  one might think that 
$(\lie_{\xi_n}\bX)_{\hat a\hat b\hat c\hat d}$
 contains the derivative of $\bX$ which makes it hopeless to evaluate, but in fact it is not.
Thanks to the $SL(2,\bR)\times U(1)$ symmetry of the background,
we have 
\begin{equation}
\partial_r(\bX_{\hat a\hat b | \hat c\hat d})=\partial_\varphi(\bX_{\hat a\hat b | \hat c\hat d})=0,
\end{equation} as is shown in Appendix~\ref{tricks}.
Then one finds
\begin{equation}
(\lie_{\xi_n} \bX)_{\hat a\hat b| \hat c\hat d}= (\xi_n)_{,\hat a}^{\hat i} \bX_{\hat i\hat b | \hat c\hat d} +\cdots.
\end{equation} 
After a slightly messy calculation, one finds that 
\begin{equation}
-2\int_\Sigma (\lie_{\xi_n} \bX)_{cd} \nabla^{[c} \xi_{-n}^{d]} \bigm|_{n^3}
=-4i n^3  \int_\Sigma \left[  
k(Z_{\hat t\hat r \hat t \hat r} - Z_{\hat t\hat\varphi\hat t\hat\varphi})
-2Z_{\hat t \hat \varphi \hat r \hat \theta }  
\frac{A(\theta) A'(\theta)}{B(\theta)}
\right]\vol(\Sigma).\label{2nd-term}
\end{equation} 
Here the prime in $A'(\theta)$ stands for the derivative with respect to $\theta$.

\subsection{The third term}\label{third}
Let us discuss the contribution from the third term, \begin{equation}
-2\int_\Sigma (\lie_{\xi_{n}} \bW_c) \xi_{-n}^c \bigm|_{n^3}.
\end{equation}  
To get something proportional to $n^3$ from the first term, we need to provide
$n$ from $\xi_{-n}^c$ and $n^2$ from $\lie_{\xi_n} \bW_c$.
Thus the index $c$ needs to be the $\hat r$ direction, and moreover
the Lie derivative needs to  provide $\xi^{\hat r}_{,\hat \varphi}$.
From the formula of the Lie derivative in the vierbein basis \eqref{lie-formula},
we find we need to have $\bW_{\hat r|\hat r\hat \theta}$ 
to use $\xi^{\hat r}_{,\hat\varphi}$. Therefore we have \begin{equation}
-2\int_\Sigma (\lie_{\xi_{n}} \bW_c) \xi_{-n}^c \bigm|_{n^3}
=
-2\int_\Sigma (\xi_n)^{\hat r}_{,\hat \varphi} \bW_{\hat r|\hat\theta\hat r} \xi_{-n}^{\hat r} e^{\hat\theta}e^{\hat\varphi}.
\end{equation} Now from \eqref{defW} we have \begin{equation}
\bW_{\hat r|\hat\theta \hat r}=-4\nabla^{\hat d} Z_{\hat t \hat\varphi \hat r \hat d}.\label{Ws=0}
\end{equation} Expanding the covariant derivative in terms of
ordinary derivatives plus spin connection terms, one finds\begin{multline}
-2\int_\Sigma \lie_{\xi_n} \bW_c \,\xi^{c}_{-n} \bigm|_{n^3}
=-2\int d\theta d\varphi \Big[ 4i A^2\partial_\theta Z_{\hat t\hat\varphi \hat r \hat \theta} \\
+2ikB(Z_{\hat t\hat r\hat t\hat r}+Z_{\hat r\hat\varphi\hat r\hat\varphi})
-4iAA'(Z_{\hat t\hat r\hat\theta\hat\varphi}-Z_{\hat t\hat\varphi\hat r\hat\theta})
+4i\frac{A^2B'}B (Z_{\hat t\hat\theta\hat r\hat\varphi} + Z_{\hat t\hat\varphi\hat r\hat \theta}) \Big].
\end{multline}
As detailed in Appendix~\ref{tricks}, the $SL(2,\bR)\times U(1)$ invariance of the metric \eqref{metric}
implies 
\begin{equation}
Z_{\hat r\hat\varphi\hat r\hat\varphi}=-Z_{\hat t\hat\varphi\hat t\hat\varphi},
\end{equation} and also using the $t$-$\varphi$ reflection symmetry
one can show \begin{equation}
Z_{\hat t\hat r\hat \theta\hat\varphi}=-2Z_{\hat t\hat\varphi\hat r\hat \theta}
,\qquad
Z_{\hat t\hat\theta\hat r\hat\varphi}=-Z_{\hat t\hat\varphi\hat r\hat \theta}.
\end{equation} 
Combining them and partially integrating once, we find
\begin{equation}
-2\int_\Sigma \lie_{\xi_n} \bW_c \,\xi^{c}_{-n} \bigm|_{n^3} 
=-2n^3\int_\Sigma d\theta d\varphi\left[
2ikB(Z_{\hat t\hat r\hat t\hat r} -  Z_{\hat t\hat\varphi\hat t\hat\varphi})+
4i AA'Z_{\hat t\hat \varphi\hat r\hat\theta}
\right].\label{3rd-term}
\end{equation}
Combining with the second term \eqref{2nd-term}, one finds \begin{equation}
-2\int_\Sigma \left[ (\lie_{\xi_n} \bX)_{cd} \nabla^{[c} \xi_{-n}^{d]}
 + \lie_{\xi_n} \bW_c \,\xi^{c}_{-n} \right]\Bigm|_{n^3}  
 =-8i k n^3\int_\Sigma d\theta d\varphi B(\theta)(Z_{\hat t\hat r\hat t\hat r} -  Z_{\hat t\hat\varphi\hat t\hat\varphi}). \label{remainder}
\end{equation}
Note that \begin{equation}
Z_{\hat t\hat r \hat t \hat r} - Z_{\hat t\hat\varphi\hat t\hat\varphi}
\end{equation} is zero for the Einstein-Hilbert  Lagrangian, because \begin{equation}
Z_{abcd}=\frac{1}{16\pi G_N} (g_{ac}g_{bd}-g_{ad}g_{bc}),
\end{equation}
but it is not zero in general. For example, it is nonzero when
$Z_{abcd}$ contains a term proportional to $R_{abcd}$, 
which is the case when there is a term $\alpha R_{abcd} R^{abcd}$
in the Lagrangian. Therefore we conclude that the charge as defined by Iyer-Wald, \eqref{kIW}
does not reproduce the Iyer-Wald entropy. 

\subsection{The term $\bE$}
We now show the charge advocated  in \citen{Barnich:2001jy,Barnich:2007bf,Compere:2007az},
 \eqref{kBC}, indeed reproduces the Iyer-Wald entropy.
The difference of $\bk^{IW}$ and $\bk^{inv}$  is given by the $\bE$-term \eqref{Eterm}.
Combining with \eqref{formula},
 one has \begin{equation}
\int_\Sigma \bk_{\xi_n}^{inv} [\lie_{\xi_{-n}}\phi;\bar\phi]  \bigm|_{n^3}
= -2\int_\Sigma \lie_{\xi_n}\bQ_{\xi_{-n}}\bigm|_{n^3}
 - \int_\Sigma \bE[\lie_{\xi_{n}}\phi,\lie_{\xi_{-n}}\phi; \bar\phi] \bigm|_{n^3}.
\end{equation}
We can easily see that \begin{equation}
\int_\Sigma \bE[\lie_{\xi_{n}}\phi,\lie_{\xi_{-n}}\phi; \bar\phi] \bigm|_{n^3}
\end{equation} gives \begin{equation}
=8ikn^3\int_\Sigma d\theta d\varphi B(\theta)(Z_{\hat t\hat r \hat t\hat r} - Z_{\hat t\hat\varphi\hat t\hat\varphi}),
\end{equation} which perfectly cancels \eqref{remainder}. 

Therefore we have \begin{equation}
\int_\Sigma \bk_{\xi_n}^{inv} [\lie_{\xi_{-n}}\phi; \bar\phi]  \bigm|_{n^3}
=-i n^3 k \int_\Sigma Z_{abcd} \epsilon^{ab}\epsilon^{cd}\vol(\Sigma).
\end{equation} Using the Cardy formula at the Frolov-Thorne temperature \begin{equation}
T_{FT}=\frac{1}{2\pi k},
\end{equation} we find that the central charge of the asymptotic Virasoro algebra exactly reproduces
the Iyer-Wald entropy.
We conclude that the central charge
of the asymptotic Virasoro symmetry reproduces the entropy
{\em if and only if} one includes the correction terms advocated in \citen{Barnich:2007bf,Compere:2007az} following the definitions of \citen{Barnich:2001jy}.

\subsection{Lagrangians with derivatives of Riemann tensor}\label{later}
Let us see what needs to be changed when we deal with Lagrangians with derivatives of Riemann tensor.
From the form of $\bQ_\xi$ in \eqref{Qs} for such a Lagrangian,
we see that the only possible change in the central charge is that $\bW_c$ in \eqref{formula} 
becomes \begin{equation}
\bW_c=\sum_s \bW^{(s)}_c,
\end{equation}
where $\bW^{(0)}_c$ is given in \eqref{defW} and
\begin{equation}
\bW_{k|c_3 c_4}^{(s)} = -2 \Big(Y_{kab}^{(s)}+Y_{abk}^{(s)}+Y_{akb}^{(s)}+\frac{s-1}2U_{kab}^{(s)}\Big) \epsilon^{ab}{}_{c_3c_4},
\end{equation} for $s\ge 1$, where \begin{align}
Y_{kab}^{(s)}&=Z_{klcd|ae_1\cdots e_{s-1}} \bR_b{}^{lcd}{}_{|e_1\cdots e_{s-1}},\\
U_{kab}^{(s)}&=Z_{mlcd|ka e_1\cdots e_{s-2}} \bR^{mlcd}{}_{be_1\cdots e_{s-2}}.
\end{align}
Therefore, what we need to show is that the contribution \begin{equation}
-2\int_\Sigma (\lie_{\xi_{n}} \bW_c^{(s)}) \xi_{-n}^c \bigm|_{n^3}.
\end{equation} in \eqref{formula} vanishes for each $s\ge 1$.

In the rest of this subsection we drop $\hat{} $ and $^{(s)}$ for the sake of brevity.
As  in Sec.~\ref{third}, only the component  $\bW_{ r| \theta  r}$
contributes to the $\cO(n^3)$ term,
which is \begin{equation}
\bW_{r|\theta r}=2\Big(Y_{rt\varphi}-Y_{r\varphi t} + Y_{t\varphi r}-Y_{\varphi t r} +Y_{tr \varphi }-Y_{\varphi r t} +\frac{s-1}{2}(U_{r t\varphi} - U_{r\varphi t})\Big).
\end{equation} Now, using the $SL(2,\bR)\times U(1)$ invariance and the $t$--$\varphi$ reflection
as detailed in Appendix~\ref{tricks},
we have $Y_{rt\varphi}=-Y_{tr\varphi}$, and  their cyclic permutations.
Also, because $U_{kab}$ is symmetric in $k$ and $a$, one has $U_{rt\varphi}=0$.
 Thus we have \begin{equation}
\bW_{r|\theta r}=-4Y_{r\varphi t}-(s-1)U_{r\varphi t}.
\end{equation}

For $s=1$, we only need to show $Y_{r\varphi t}=0$. Expanding $Y$,
we have \begin{equation}
Y_{t\varphi r}=-2( \bR_{t\theta t\theta} Z_{t\theta r\theta | \varphi}+
\bR_{t\varphi t\varphi} Z_{t\varphi r\varphi | \varphi}),
\end{equation} where we used the $SL(2,\bR)\times U(1)$ invariance of 
$\bR_{abcd}$ and $Z_{abcd|e}$.  
Now $Z_{r\theta t\theta | \varphi}=Z_{t\theta r\theta | \varphi}$
because of the symmetry of the Riemann tensor, 
but under the $t$-$\varphi$ reflection
we have $Z_{r\theta t\theta | \varphi}=-Z_{t\theta r\theta | \varphi} $
as argued in Appendix~\ref{tricks}. Thus we have $Z_{r\theta t\theta|\varphi}=0$,
and similarly we can show $Z_{t\varphi r\varphi | \varphi}=0$. We conclude 
$\bW^{(1)}_{r\varphi t}=-4Y^{(1)}_{r\varphi t}=0$.

For $s>1$ we have not found a pencil-and-paper proof of the vanishing of $\bW^{(s)}_{r|\theta r}$,
but we implemented the symmetry properties detailed in Appendix~\ref{tricks} in Mathematica
and checked that  identically $\bW^{(s)}_{r|\theta r}$ 
vanishes up to $s=35$.\footnote{It took about two hours to perform this calculation on a 3GHz machine.
The Mathematica file can be obtained upon request to Y.~T. 
}
Therefore we strongly believe that  it vanishes for all $s\ge 1$.
Our conclusion is then that the central charges of the boundary Virasoro symmetry
correctly reproduces the Iyer-Wald entropy of the black hole for arbitrary diffeomorphism-invariant Lagrangian
constructed solely from the metric, when we use the asymptotic charges defined in \citen{Barnich:2001jy,Barnich:2007bf,Compere:2007az}.

\section{Summary and discussion}\label{summary}
In this paper, we studied the Dirac bracket of the asymptotic Virasoro symmetry acting on the near-horizon geometry of the 4d extremal black holes in gravity theories with higher-derivative corrections.
We first determined  the explicit form of the asymptotic charges in the presence of higher-derivative corrections in the Lagrangian, and then used it to evaluate the central charge. 
After a laborious calculation, we found that the entropy formula of Iyer-Wald is perfectly reproduced, once one carefully includes the boundary term in the asymptotic charge advocated in \citen{Barnich:2001jy,Barnich:2007bf,Compere:2007az}.
This result gives us reassurance that it is not just a numerical coincidence owing to the simple form of the Einstein-Hilbert Lagrangian that the Bekenstein-Hawking entropy and the entropy determined from the asymptotic Virasoro symmetry agreed in the original paper\cite{Guica:2008mu} and in the generalizations. 
In view of our findings, there should indeed be a Virasoro algebra acting on the microstates of the four-dimensional extremal black hole, which accounts for its entropy.

If we remember that the Cardy formula is 
valid in the high temperature limit,
then it is natural to ask how the corrections to the entropy 
from the higher-derivative terms will be distinguished from
the corrections to the Cardy formula.
For the black hole with several charges, for example \citen{Azeyanagi:2008kb}, 
we can think the temperature $T_{FT}=1/2\pi k$ as
an independent parameter and take the high temperature limit
with the other chosen parameters including the Planck length $l_p$ fixed. 
Then the Cardy formula is expected to be valid for
the leading order in k,
and it should be matched with the leading order
of the Iyer-Wald entropy, 
which will include many higher-derivative corrections.
However, our result is too much better than expected:
we found that the Cardy formula exactly reproduced the Iyer-Wald entropy.
Indeed, this mysterious accuracy of the Cardy formula has already been observed for the case 
without higher-derivative terms, see \citen{Strominger:1997eq,Guica:2008mu}. It would be interesting to investigate reason for it.

There are a few straightforward but calculationally intense directions to extend our work presented here. 
Namely, in this paper we only studied asymptotic Virasoro symmetry of the 4d extremal black holes in a theory whose only dynamical fields are the metric and its auxiliary fields. 
Then it would be natural to try to extend it to black holes in higher dimensions, to theories with scalars and vectors with higher-derivative corrections, 
and to theories with gravitational Chern-Simons and Green-Schwarz terms.  We leave these endeavors to daring individuals with plenty of time to spare.

The most pressing issue is, unarguably, the question of the nature of the Virasoro symmetry acting on the  microstates, of which our work unfortunately does not have much to tell.  For the standard AdS/CFT correspondence, the CFT on which the conformal symmetry acts can be thought to live on the boundary of the spacetime. Naively, one would say that in the case of extremal rotating black holes, this boundary CFT lives on
one of the two time lines being the boundary of the AdS$_2$ part of the metric. 
For a specific example of D1-D5-P black holes, three of the authors showed that this Virasoro symmetry is a part of the conformal symmetry of the CFT on the brane system\cite{Azeyanagi:2008dk}. In the usual AdS/CFT correspondence, 
we have the prescription\cite{Gubser:1998bc,Witten:1998qj} which extract the information of the CFTs without referring to the string theory embedding, given the bulk gravity solution. It would be preferable if we have an analogue of that in the extremal black hole/CFT correspondence, and we would like to come back to this question in the future.

\section*{Note Added}

During the completion of this work, the paper \citen{Krishnan:2009tj} appeared in which it was shown that the formalism of \citen{Barnich:2001jy,Barnich:2007bf,Compere:2007az} applied to the Gauss-Bonnet theory formulated using the metric only cannot reproduce the Iyer-Wald entropy.
Here, we proved that using auxiliary fields to take into account the higher-derivative corrections to the Einstein-Hilbert Lagrangian, the formalism of \citen{Barnich:2001jy,Barnich:2007bf,Compere:2007az} reproduces the correct Iyer-Wald entropy. One consequence of these two computations is that the formalism of \citen{Barnich:2001jy,Barnich:2007bf,Compere:2007az} is not invariant under field redefinitions. 
In view of the cohomological results of \citen{Barnich:2001jy}, this ambiguity can appear only in the asymptotic context and when certain asymptotic linearity constraints are not obeyed. It has been acknowledged that boundary terms in the action should be taken into account \cite{Regge:1974zd,Hawking:1995fd}. 
Adding supplementary terms to a well-defined variational principle amount to deforming the boundary conditions\cite{Breitenlohner:1982jf,Witten:2001ua,Marolf:2006nd} and modifying the symplectic structure of the theory through its coupling to the boundary dynamics \cite{Compere:2008us}. It would be interesting to understand how these boundary effects would contribute in relation to the work of \citen{Krishnan:2009tj}.

\section*{Acknowledgment}
The authors thank helpful discussions with K. Murata and T. Nishioka. G.~C. also thanks S. Detournay, G. Horowitz, D. Marolf, M. Roberts and A. Strominger for fruitful discussions and G. Barnich, C. Krishnan and T. Hartman for interesting correspondences. T.~A. is supported by the Japan 
Society for the Promotion of Science (JSPS).
The work of G.~C. is supported in part 
by the US National Science Foundation under Grant No. PHY05-55669, and 
by funds from the University of California. 
Y.~T. is supported in part by the NSF grant PHY-0503584, and in part by the Marvin L. 
Goldberger membership at the Institute for Advanced Study. 
S.~T. is partly supported 
by the Japan Ministry of Education, Culture, Sports, Science and 
Technology (MEXT).
A part of this work was supported by the Grant-in-Aid for the 
Global COE program ``The Next Generation of Physics, Spun from 
Universality and Emergence'' from the MEXT.

\appendix

\section{Integrability and finiteness of charges}\label{integrability}

In this appendix, we investigate the integrability 
and finiteness
of the charges $\bk^{IW}$ and $\bk^{inv}$ defined in \eqref{kIW} and \eqref{kBC}. When considering general higher-derivative corrections, it is difficult to show the integrability systematically. This can be understood from the fact that we have to solve the equations of motion and this is impossible without an explicit expression for the Lagrangian. Therefore, in this paper, we only show the integrability for the case of Gauss-Bonnet gravity
\begin{align}
  \bL = \star \Big(\frac{1}{16\pi G_N}R + \alpha {L^{GB}} \Big),
\quad
L^{GB} = R_{abcd}R^{abcd} - 4R_{ab}R^{ab} + R^2,
\end{align}
and we will further limit ourselves to show integrability only around the background $\bar g$ given in \eqref{metric}.  In Gauss-Bonnet theory, the equations of motion are not deformed with respect to those of Einstein gravity. The relevant extremal black hole geometry is thus the near-horizon extremal Kerr geometry \cite{Bardeen:1999px} which takes the form of \eqref{metric} with 
\begin{equation}
A(\theta) = a \sqrt{\frac{1+\cos^2\vartheta}2},
\quad B(\theta) = a\sin\vartheta\sqrt{\frac2{1+\cos^2\vartheta}},
\end{equation} where \begin{equation}
d\vartheta=A(\theta)d\theta.
\end{equation}
The prefactor $a$ controls the mass and the angular momentum.

For the Gauss-Bonnet theory, we can just follow the appendix of \citen{Guica:2008mu} and solve the constraint condition $G^t_{a}=0$ for the metric $\bar g + \delta_{1}g$ at leading order,
where $G_{ab}$ is the Einstein tensor and $\delta_1 g$ obeys the boundary condition \eqref{bc}. We get
\begin{subequations}
\label{constraint}
\begin{align}
  \delta_1 g_{tt}&=r^2(A(\theta)^2-k^2B(\theta)^2)f^{(1)}(t,\varphi) + o(r^2),\\
  \delta_1 g_{\varphi\varphi}&=B(\theta)^2f^{(1)}(t,\varphi) + o(1),\\
  \delta_1 g_{r\varphi} &= -\frac{A(\theta)^2}{2r}\frac{\partial}{\partial\varphi}f^{(1)}(t,\varphi) + o(1/r).
\end{align}
\end{subequations}
We also define $f^{(2)}$ in the same way for another metric perturbation $\delta_{2}g$. 
Now, we can define the perturbation of the auxiliary field $Z^{abcd}$ 
around $\bar g$
using the equations of motion as
\begin{equation}
Z^{abcd}[\bar g +\delta_1 g] = \frac{\partial \cL}{\partial  R_{abcd}}  = Z^{abcd}[\bar g] + Z_{(1)}^{abcd}[\delta_1 g; \bar g] + \cO\left((\delta_1 g)^2\right). 
\end{equation}
The integrability condition for $\bk_\xi[\delta g;g]$,
in an infinitesimal neighborhood of a general background $g$, reads as
\begin{align}
\int_{\Sigma} \left( \delta \bk_{\xi}\right) [\delta_1 g,\delta_2 g; g] = 0,
\end{align}
where the fields $Z$ and $\delta Z$ have been replaced by their on-shell values in terms of $g$ and $\delta g$. Equivalently, one has to show that 
\begin{align}
\left(\delta \bk_{\xi}\right)[\delta_1 g,\delta_2 g ; g]
  =  \bk_{\xi}[\delta_1 g ; g + \delta_2 g] + \bk_{\xi}[\delta_2 g ; g]
   - \bk_{\xi}[\delta_2 g ; g+\delta_1 g] - \bk_{\xi}[\delta_1 g ; g]\label{defEB}
\end{align}
is zero for $r \rightarrow \infty$ up to boundary terms and non-linear terms in $\delta_1 g,\delta_2  g$. Now, we showed by using Maple that
\begin{align}
\left(\delta \bk^{IW}_{\xi}\right)[\delta_1 g, \delta_2 g ; \bar g] = 0,
\end{align}
at leading order in $\delta_1 g$, $\delta_2 g$ and for $r \rightarrow \infty$, when we require \eqref{constraint} for $\delta_1 g$ and $\delta_2 g$. Let us define the analogue of \eqref{defEB} for the $\bE$ term by replacing 
$\delta\bk_{\xi}$
by
$\delta\bE[\lie_\xi g]$
on the left-hand side and all occurrences of $\bk_{\xi}[\delta s_1 ; s_2]$ by $\bE[\lie_\xi s_2,\delta s_1; s_2]$ on the right-hand side. We then find using Maple that under the same conditions,
\begin{align}
  & \left( \delta \bE[\lie_\xi g]\right) [\delta_1 g, \delta_2 g ; \bar{g}] \nonumber\\
  &\;
    = 2\alpha k\frac{B''(\theta)A(\theta) - B(\theta)A''(\theta)}{A(\theta)}
    \left(
       f^{(1)} \frac{\partial^2f^{(2)}}{\partial\varphi^2}
     - \frac{\partial^2f^{(1)}}{\partial \varphi^2}f^{(2)}
     - in f^{(1)}\frac{\partial f^{(2)}}{\partial\varphi}
     + in \frac{\partial f^{(1)} }{\partial\varphi}f^{(2)}
    \right)
    e^{-in\varphi}.
\end{align}
Contrary to the case of Einstein gravity considered in \citen{Guica:2008mu}, this does not vanish locally. However, by partial integral for $\varphi$, we can easily show that
\begin{align}
  \int_{\Sigma} \left( \delta \bE[\lie_\xi g]\right) [\delta_1 g, \delta_2 g ;\bar g] = 0,
\end{align}
and of course it leads to
\begin{align}
  \int_{\Sigma} \left( \delta \bk^{inv}_{\xi}\right) [\delta_1 g,\delta_2 g ; \bar g]  = 0.
\end{align}
Therefore, we have shown that both of $\bk^{IW}_{\xi_n}$ and $\bk^{inv}_{\xi_n}$ are integrable in infinitesimal neighborhood around the metric \eqref{metric} in Gauss-Bonnet gravity. 
To show the integrability fully, we must consider the fluctuation 
around any metric satisfying the given boundary condition and show 
the integrability  but such a proof is lacking.

The finiteness of the charges corresponding to the Virasoro generators can be shown in general along the following lines. Let us consider 
a tensor ${T^{a_1a_2\dots}}_{b_1b_2\dots}$ which is made of
$\bar{g}_{ab}$, $\bar{g}^{ab}$, $\delta g_{ab}$, $\xi_n^{a}$
and their derivatives.
In particular,
\eqref{Thetas}, \eqref{Qs} and \eqref{Eterm} satisfy this condition.
From \eqref{metric}, \eqref{bc} and \eqref{Vf},
each component of this tensor behaves as
${T^{a_1a_2\dots}}_{b_1b_2\dots}=\cO(r^l)$ at most,
where 
\begin{align}
  l = - (\#\;\text{of}\;t\;\text{in}\;a_i\text{'s})
+ (\#\;\text{of}\;t\;\text{in}\;b_i\text{'s})
+ (\#\;\text{of}\;r\;\text{in}\;a_i\text{'s})
- (\#\;\text{of}\;r\;\text{in}\;b_i\text{'s}).
\end{align}
Since $\bm{k}_{\xi_n}^{IW}$ and $\bm{k}_{\xi_n}^{inv}$
consist of \eqref{Thetas}, \eqref{Qs} and \eqref{Eterm},
the $tr$ and $rt$ components of
both $\bm{k}_{\xi_n}^{IW}$ and $\bm{k}_{\xi_n}^{inv}$
behave as $\cO(1)$ at most.
Therefore the corresponding charges $H_n^{IW}$ and $H_n^{inv}$
are all finite.

Next let us consider
$\bm{k}_{\partial_t}^{IW}$ and $\bm{k}_{\partial_t}^{inv}$,
which are made of $\bar{g}_{ab}$, $\bar{g}^{ab}$, $\delta g_{ab}$,
their derivatives and $\partial_t$.
For some terms in
$\bm{k}_{\partial_t}^{IW}$ and $\bm{k}_{\partial_t}^{inv}$,
one of $t$'s in the upper indices has its origin
in $\partial_t$, which contribute as
$\cO(1)$, instead of $\cO(1/r)$.
Thus it follows that
$\delta H_{\partial_t}^{IW}$ and $\delta H_{\partial_t}^{inv}$
can diverge from this order counting of $r$. This divergence is removed once we impose the Dirac constraint $H_{\partial_t}=0$.

\section{On the Frolov-Thorne temperature}\label{FT}

In this appendix we show, under a mild assumption,
that $k$ which appears in the metric \eqref{metric}
gives the inverse Frolov-Thorne temperature \begin{equation}
T_\text{FT}=\frac1{2\pi k},
\end{equation}
even in the presence of the higher-derivative terms.
In other words,  there is no correction to the Frolov-Thorne temperature from the higher-derivative terms in the Lagrangian. It is in a sense expected: the Hawking temperature arises from the analysis of free fields on the curved background, and thus depends on the metric but not on the equations of motion which the metric solves. The Frolov-Thorne temperature should also be encoded in the metric.

The fact that there is no correction to the Frolov-Thorne temperature
coming from the matter fields has already been stated in \citen{Chow:2008dp},
see their argument leading to their (2.9). 
Here we develop their argument in detail.
We will make several assumptions in the course, 
which we try to make  as manifest as possible. 
These assumptions seem natural to us; at least they are rather qualitative.
The crucial fact is that we do not use any equation of motion, 
so the argument  should apply to generic Lagrangians,
even with higher-derivative terms.

\subsection{Non-extremal black hole and the temperature}
We suppose that there is a family of 4d rotating black hole solutions whose metric is \begin{equation}
ds^2=g_{rr} dr^2+g_{\theta\theta} d\theta^2+ a (a_t dt- a_\varphi d\varphi)^2- b ( b_t dt- b_\varphi d\varphi)^2.
\end{equation}
Here $g_{rr},g_{\theta\theta},a,a_t,a_\varphi,b,b_t,b_\varphi$ are all functions of $r$,$\theta$,
the ADM mass $M$ and the angular momentum $J$, and {\em assume they are smooth
across the horizon with respect to $r$, $M$ and $J$}.
For $g_{rr}$, we require the smoothness of $1/g_{rr}$.
This ansatz is a big assumption but is rather qualitative,
and is known to be satisfied in many examples.

We assume that the metric asymptotes to the flat space or to the AdS space so that 
the first law of the black hole is guaranteed. 
The asymptotic time translation is $\partial_t$ and the rotation is $\partial_\varphi$.

We assume that the  horizon is at $r=r_H$ which is a function of $J$ and $M$.
We write the horizon generating Killing vector as $\xi = \partial_t + \Omega_H \partial_\varphi$,
where $\Omega_H$ is the angular velocity of the horizon, which appears in the first law.
We assume,  for generic values of $M$ and $J$, \begin{equation}
g_{rr} \sim \cO(1/\delta r), \qquad
b \sim \cO(\delta r), \quad
a \sim \cO(1), \quad
a_t  -a_\varphi \Omega_H \sim \cO(\delta r).
\end{equation} close to the horizon, $\delta r=r-r_H$.

The temperature is given by $\kappa/(2\pi)$, where
the surface gravity 
\begin{equation}
\kappa= \sqrt{- \frac12 g^{ac}g^{bd}\nabla_a \xi_b \nabla_c \xi_d }
\end{equation}
 is evaluated at the horizon. To evaluate it, it is convenient to use the fact \begin{equation}
 d\xi = \nabla_a \xi_b dx^a \wedge dx^b
\end{equation} for a Killing vector $\xi$.
Here $d\xi$ is the exterior derivative of the one-form $\xi=g_{ij}\xi^i dx^j$.
We have \begin{equation}
\xi = a(a_t - a_\varphi \Omega_H)(a_t dt+ a_\varphi d\varphi) +
b(b_t - b_\varphi \Omega_H)(b_t dt+ b_\varphi d\varphi).
\end{equation}  
Once one rewrites it using the vierbein basis $\sqrt{g_{rr}} dr$,
$\sqrt{g_{\theta\theta}} d\theta$, etc., one finds that most of the term
goes to zero at $r=r_H$ because \begin{equation}
a(a_t-a_\varphi \Omega_H) \sim \cO(\delta r), \qquad
b(b_t-b_\varphi \Omega_H) \sim \cO(\delta r),
\end{equation} and that the only term which contributes to $d\xi$ on the horizon
 is \begin{equation}
\frac{\partial}{\partial r}\left[
b(b_t -b_\varphi\Omega_H)
\right] dr\wedge (b_\varphi dt + d\varphi).
\end{equation}
Therefore \begin{equation}
T_H=\frac{\kappa}{2\pi} =\frac1{4\pi}
\frac{(b_t-b_\varphi \Omega_H)}{ \sqrt{b \cdot g_{rr}} }\Bigm|_{r=r_H}
\frac{\partial b }{\partial r}\Bigm|_{r=r_H}.\label{TH}
\end{equation}

\subsection{$k$ as defined by the extremal metric}
Now suppose at $M=M(J)$ the black hole becomes extremal, i.e. \begin{equation}
1/g_{rr} = \delta r^2/G + \cdots,\qquad 
b= B \delta r^2 + \cdots,
\end{equation} where $G$ and $B$ are functions of $\theta$ only.

We perform the coordinate change \begin{equation}
\delta r=\lambda \tilde\rho,\qquad
t=\tilde\tau/\lambda,\qquad
\varphi=\tilde\varphi + \Omega_H \tilde\tau/\lambda,
\end{equation} and take the limit $\lambda\to 0$. The metric becomes \begin{equation}
ds^2=G \frac{d\tilde\rho^2}{\tilde \rho^2} +g_{\theta\theta} d\theta^2
+(a a_\varphi^2 )|_{r_H} \Big( -\frac{\partial (a_t/a_\varphi)}{\partial r} \Big|_{r_H} \tilde\rho d\tilde \tau +   d\tilde\varphi \Big)^2 + (b_t - \Omega_H b_\varphi)^2|_{r_H} B (\tilde\rho d\tilde \tau)^2.
\end{equation}

Now in Kunduri-Lucietti-Reall \cite{Kunduri:2007vf}, it is shown that
there is a constant $c$  such  that \begin{equation}
 G(\theta) = c^2 (b_t(\theta) - \Omega_H b_\varphi(\theta))^2|_{r_H} B(\theta),
\end{equation}  and there is a symmetry enhancement  to $SL(2,\bR)$.
We make another change of variables \begin{equation}
 \tilde \rho=\rho,\qquad
  \tilde \tau = c \tau,
\end{equation} to arrive at \begin{equation}
ds^2=G(\theta) (\frac{d\rho^2}{\rho^2} + \rho^2 d\tau^2) + 
(a a_\varphi^2 )|_{r_H} ( k_{m} \rho d \tau +   d\tilde\varphi )^2,
\end{equation} where \begin{eqnarray}
k_m &=& -c \frac{\partial (a_t/a_\varphi)}{\partial r} \Big|_{r_H}\\
&=& -\left(
\frac{(b_t-b_\varphi \Omega_H)}{ \sqrt{b \cdot g_{rr}} }\Bigm|_{r=r_H}
\frac12\frac{\partial^2b}{\partial r^2} \Bigm|_{r=r_H}
\right) ^{-1} \frac{\partial (a_t/a_\varphi)}{\partial r} \Big|_{r_H}. \label{km}
\end{eqnarray}
The subscript $m$ emphasizes that this is $k$ as defined by the {\em metric.}
Note that the factor $c$ in (\ref{km}) is  quite similar in appearance to the expression of $T_H$, see \eqref{TH}.

\subsection{Frolov-Thorne temperature as defined from the limit of the first law}

Now let us perform the limiting of the first law: we start from \begin{equation}
T_HdS= dM -\Omega_H dJ = \frac{\d M}{\d\epsilon} d\epsilon +
\left(\frac{\d M}{\d J}-\Omega_H\right)dJ, \label{firstlaw} 
\end{equation}
where we changed the variables from $(M,J)$ to $(\epsilon,J)$ where $\epsilon$ measures the 
deviation from extremality. Here, $T_H$ is given by $\kappa/2\pi$ and $\Omega_H$ is what 
appears in $\xi=\partial_t + \Omega_H \partial_\varphi$. These relations are known not to be corrected by the higher derivatives, etc.

Let us assume the expansion of the form \begin{eqnarray}
T_H &=& \epsilon T_H'  + \cO(\epsilon^2), \\
M &=& M(J) + \epsilon M'(J) + \cO(\epsilon^2), \\
\Omega_H &=& \Omega_H(J) + \epsilon \Omega_H'(J) + \cO(\epsilon^2). 
\end{eqnarray} Here ${}'$ stands for the derivative with respect to $\epsilon$, not to $J$.
We substitute these expansions into both sides of 
(\ref{firstlaw}) and compare them order by order. 
By considering terms at order $\epsilon^0$, we obtain 
\begin{eqnarray}
 M'(J) =0, \qquad \Omega_H(J)= \frac{\partial M(J)}{\partial J}. 
\end{eqnarray} 
At order $\epsilon^1$, we then find 
\begin{eqnarray}
T_H^\prime dS = M''(J) d\epsilon - \Omega_H' dJ ,
\end{eqnarray} 
which implies that at extremality, 
\begin{equation}
 T_\text{FT} dS(J)= dJ \qquad \hbox{where}\qquad  
 T_\text{FT}=\frac{1}{2\pi k_{1st}}\quad \text{and}\quad  
 k_{1st}= - \frac{1}{2\pi}\frac{\partial \Omega_H/ \partial \epsilon }{\partial T_H/\partial \epsilon } \Big|_{\epsilon=0}.\label{k1st}
\end{equation}

\subsection{Frolov-Thorne temperature and $k$}
Now let us define \begin{eqnarray}
T&=&\frac1{4\pi}
\frac{(b_t-b_\varphi \Omega_H)}{ \sqrt{b \cdot g_{rr}} }
\frac{\partial b }{\partial r}, \\
\Omega &=& a_t / a_\varphi, 
\end{eqnarray} which are functions of $r,\theta$ and $\epsilon,J$.
They become the Hawking temperature $T_H$ and the angular velocity $\Omega_H$
when evaluated at $r=r_H$.
Then  the formula for $k_{1st}$, (\ref{k1st}) can be rewritten as \begin{equation}
k_{1st}=-\frac{1}{2\pi}\frac{\partial \Omega(r=r_H)/ \partial \epsilon }{\partial T(r=r_H)/\partial \epsilon } \Big|_{\epsilon=0}, 
\end{equation}
whereas the formula for $k_m$, (\ref{km}) can be rewritten as \begin{equation}
k_{m}=-\frac{1}{2\pi}\frac{\partial \Omega(r)/ \partial r }{\partial T(r)/\partial r } \Big|_{r=r_H}.
\end{equation} 
The final trick is to use $r_H$ itself as the extremality parameter $\epsilon$ \begin{equation}
\epsilon = r_H(M,J)-r_H^\text{extremal}(J),
\end{equation} which shows $k_{1st}=k_m$. Thus we conclude \begin{equation}
T_\text{FT}=\frac{1}{2\pi k_m}.
\end{equation}

\section{Conventions on variational calculus}\label{homotopy}
Here we summarize our conventions  used in the variational calculus.
We basically follow the conventions in \citen{Barnich:2007bf,Compere:2007az},
but change the notations to match those by the Iyer-Wald school.

We consider a spacetime $\cM$ with coordinates $x^a$,
on which fields $\phi^i$ and their derivatives $\phi_{,a}^i$, \dots treated as independent fields live. $\phi^i$ stands for all the fields including the metric.
We consider differential forms which not only include $dx^a$,
but also $\delta\phi^i$. The idea is that the one-form $dx^a$ is the mathematically
formalized version of physicist's idea of infinitesimal distance on $\cM$.
The field variation can also be formalized, as the one-forms $\delta\phi^i$.
We have differential forms generated by 
\begin{equation}
dx^a,dx^b,\ldots;\qquad
\delta\phi^i,\  \delta\phi^i_{,a},\  \delta\phi^i_{,ab},\ldots,
\end{equation}
where $\delta\phi^i_{,a}=\partial_a \delta\phi^i$, etc.
These all anti-commute with each other, since they are one-forms.
A form with $p$ $dx^i$'s and $q$ $\delta\phi^i_I$'s is called a $(p,q)$-form, where 
$I,J$ stand for  multi-indices.
Correspondingly there are two operations 
\begin{align}
d(\cdots)&=dx^a \wedge \partial_a (\cdots),\\
\delta(\cdots)&\equiv \delta\phi^i_{,I} \wedge \frac{\partial}{\partial \phi^i_{,I}}(\cdots)\\
&\equiv \left(\delta\phi^i \wedge \frac{\partial}{\partial \phi^i}
+\delta\phi^i_{,a} \wedge \frac{\partial}{\partial \phi^i_{,a}}
+\delta\phi^i_{,ab} \wedge \frac{\partial}{\partial \phi^i_{,ab}}+\cdots
\right) (\cdots).
\end{align}
$d$ is our usual total differential, and $\delta$ is our usual field variation.
They are called $d_H$ and $d_V$ respectively, in \citen{Barnich:2007bf,Compere:2007az}.
These two operations anti-commute, \begin{equation}
\{d,\delta\}=0.
\end{equation}

For a possible symmetry operation \begin{equation}
\phi^i \longrightarrow \phi^i + \epsilon \delta_Q \phi^i (\phi^j, \phi^j_{,a},\ldots),
\end{equation} we require
\begin{align}
\phi^i_{,a} &\longrightarrow \phi^i_{,a} 
+ \epsilon \partial_a \delta_Q \phi^i (\phi^j, \phi^j_{,a},\ldots),\\
\phi^i_{,ab} &\longrightarrow \phi^i_{,ab} 
+ \epsilon \partial_a\partial_b \delta_Q \phi^i (\phi^j, \phi^j_{,b},\ldots).
\end{align}
In the jet bundle approach, one first introduces the symbols
$\phi^i_{,ab}$ etc. as formal coordinates, and so a general vector field on the jet bundle
will {\em not} satisfy this property.  That is why there is a need to distinguish a vector field
and its prolongation in general.

We also define the interior product to be \begin{equation}
{\partial_a}\interior dx^b = \delta_a^b, \qquad 
{\partial_a} \interior  \delta\phi^i_{,bc}=0,
\end{equation} etc.
 Thus, by definition, we have \begin{equation}
 \delta_Q (\phi^i_{,a})= \partial_a \delta_Q\phi^i
\end{equation} and we define \begin{equation}
\delta_Q (\delta\phi^i) \equiv \delta(\delta_Q\phi^i).
\end{equation}

 The definition of $\partial/\partial\phi^i_{,ab}$ is \begin{align}
\frac{\partial}{\partial \phi^i_{,ab}} dx^b&=0, &
\frac{\partial}{\partial \phi^i_{,ab}} \phi^j_{,cd}
&=\delta^i_j \delta^{(a\vphantom{b}}_{c\vphantom{d}} \delta^{b)}_d,
\end{align} etc.\ Note that this includes the symmetrization factor, e.g.~$\partial \phi_{,xy}/\partial\phi_{,xy}=1/2$.
 
Higher order Euler-Lagrange derivatives are \begin{align}
\frac{\delta}{\delta \phi^i_{,I}} = \sum_{J} (-1)^J \binom{|I| + |J|}{|J|}  \partial_J\frac{\partial}{\partial \phi^i_{,IJ}},
\end{align}
where $I,J$ stand for the multi-indices; more concretely, we have equations \begin{align}
\frac{\delta}{\delta \phi^i} &=
 \frac{\partial}{\partial \phi^i} -\partial_a  \frac{\partial}{\partial\phi^i_{,a }}
+\partial_a \partial_b  \frac{\partial}{\partial\phi^i_{,a b }} - \cdots,\\
\frac{\delta}{\delta \phi^i_{,a }} &=
 \frac{\partial}{\partial \phi^i_{,a }} -2\partial_b  \frac{\partial}{\partial\phi^i_{,a b }}
+3\partial_b \partial_c \frac{\partial}{\partial\phi^i_{,a b c}} - \cdots,\\
\frac{\delta}{\delta \phi^i_{,a b }} &=
 \frac{\partial}{\partial \phi^i_{,a b }} -3\partial_c \frac{\partial}{\partial\phi^i_{,a b c}}
+6 \partial_{c}\partial_d \frac{\partial}{\partial\phi^i_{,a b cd}} - \cdots.
\end{align}

The homotopy operators are then \begin{equation}
I_{\delta \phi}^p \bomega = \sum_I \frac{|I|+1}{n-p+|I|+1} \partial_I \left[ \delta\phi^i \wedge
\frac\delta{\delta\phi^i_{,Ib }} (\partial_b  \interior \bomega) \right],
\end{equation} where $n$ is the spacetime dimension
and $\bomega$ is a $(p,q)$-form. $I_{\delta \phi}^p\bomega$ is then a $(p-1,q+1)$ form.

Explicitly, they are \begin{align}
I_{\delta \phi}^n \bomega&= 
\delta\phi^i\wedge \frac\delta{\delta\phi^i_{,a }} \partial_a \interior\bomega 
+\partial_a  \left[\delta\phi^i\wedge \frac\delta{\delta\phi^i_{,a b }} \partial_b \interior\bomega \right]+\cdots \\
&= \delta\phi^i\wedge \frac\partial{\partial\phi^i_{,a }} \partial_a \interior\bomega 
-\delta\phi^i\wedge \partial_b  \frac\partial{\partial\phi^i_{,a b }} \partial_a \interior\bomega 
+\delta\phi^i_{,a }\wedge \frac\partial{\partial\phi^i_{,a b }} \partial_b \interior\bomega +\cdots, \\
I_{\delta \phi}^{n-1} \bomega&= 
\frac12\delta\phi^i\wedge \frac\delta{\delta\phi^i_{,a }} \partial_a \interior\bomega 
+\frac23\partial_a  \left[\delta\phi^i\wedge \frac\delta{\delta\phi^i_{,a b }} \partial_b \interior\bomega \right]+\cdots \\
&= \frac12\delta\phi^i\wedge \frac\partial{\partial\phi^i_{,a }} \partial_a \interior\bomega 
-\frac13\delta\phi^i\wedge \partial_b  \frac\partial{\partial\phi^i_{,a b }} \partial_a \interior\bomega 
+\frac23\delta\phi^i_{,a }\wedge \frac\partial{\partial\phi^i_{,a b }} \partial_b \interior\bomega +\cdots.
\end{align}

In  our  paper, we deal with Lagrangians which contain
arbitrarily high derivatives of the Riemann tensor, but we introduce towers of
auxiliary fields so that the derivatives in the Lagrangian is 
of the second order at most. Then the formulae written above suffice.

When dealing with conserved charges, it is convenient to add new fields $\xi^\alpha$, $\xi^{\alpha}_{,a}$, \dots to the jet bundle. An homotopy $I_{\xi}^p$ mapping $(p,q)$-forms to $(p-1,q)$-forms can then be defined. When it acts on forms $\bomega_{\xi}$ linear in the fields $\xi^\alpha$ and $\xi^\alpha_{,a}$ only, the homotopy $I_{\xi}^p$ takes the form 
\begin{equation}
I_{\xi}^p \bomega_{\xi} = \frac{1}{n-p}\xi^\alpha \frac{\d}{\d \xi^\alpha_{,a}} \d_a \interior\bomega_{\xi}.
\end{equation}

\section{Consequences of the isometry}\label{tricks}
\subsection{Consequence of $SL(2,\bR)\times U(1)$ invariance}

Let us take a point $p$ on the extremal background \eqref{metric}, say at $r=1,t=0$ and
at fixed values of the angular coordinates $\theta,\varphi$.
Then a one-parameter subgroup of $SL(2,\bR)\times U(1)$  fixes the point.
In terms of
the Killing vectors \eqref{killing},
it is generated by \begin{equation}
\zeta_p\equiv \zeta_1 -2\zeta_3 - 2k \zeta_0.
\end{equation} 
As the vector $\zeta_p$ fixes  the point $p$,
$\zeta_p$ generates a Lorentz transformation on the tangent space $T_p$ at that point.
Studying the action of $\zeta_p$ to the vierbeine  at $p$ given in \eqref{vierbein} explicitly,
one  finds that it is just a Lorentz boost along  the $e^{\hat t}$-$e^{\hat r}$ plane at $p$:
\begin{equation}
\lie_{\zeta_p} e^{\hat t} = e^{\hat r}/r, \qquad
\lie_{\zeta_p} e^{\hat r} = e^{\hat t}/r.
\end{equation}
It means that every tensor constructed out of the metric, scalar, etc.\ is invariant under this boost. This imposes many conditions on the components of tensors.
For example, any vector component  $T_{\hat r}$ or $T_{\hat t}$ is zero
because they cannot be invariant under the boost.
To study tensors with more indices,
it is convenient to introduce $e^{\hat\pm} =e^{\hat t} \pm e^{\hat r}$. 
Then, tensors invariant under the boost need to have the same number of 
$\hat +$ and $\hat -$ indices.
Take a two index tensor $T_{ab}$ for illustration. We immediately have
\begin{equation}
T_{\hat +\hat +}=T_{\hat -\hat-}=0,
\end{equation} and the only nonzero components 
are $T_{\hat +\hat -}$ and $T_{\hat-\hat+}$.
$T_{\hat+\hat-}=\pm T_{\hat-\hat+}$ depending on the (anti)symmetry
of $T_{ab}$. Translated back to ($\hat r$, $\hat t$) basis, one finds \begin{equation}
T_{\hat t\hat t}=-T_{\hat r\hat r}, \qquad T_{\hat t\hat r}=0
\end{equation} for a symmetric tensor, and \begin{equation}
T_{\hat t\hat t}=T_{\hat r\hat r}=0, \qquad T_{\hat t\hat r}=-T_{\hat r\hat t}
\end{equation} for an antisymmetric tensor.

Another example is a mixed component $Z_{\hat r \hat t \hat r \hat\theta}$
of a four-index tensor: it has three indices of $\hat r$ or $\hat t$,
which translate to three indices of $\hat +$ or $\hat -$.
Therefore  this component is zero.

Another observation is that, if one assumes the tensors $T_{\hat a\hat b\hat c \ldots}$ 
to be invariant under $SL(2,\bR)\times U(1)$, then \begin{equation}
\partial_r T_{\cdots} =\partial_t T_{\cdots} =\partial_\varphi T_{\cdots}=0,
\label{derivatives-of-components}
\end{equation}  where $T_{\cdots}$ stands for the components in the vierbein basis.
To see this, we first observe
$\lie_{\zeta}e^{\hat i}$=0 for $\zeta_{0,1,2}$ of $G=SL(2,\bR)\times U(1)$.
Now let us consider a tensor \begin{equation}
T\equiv T_{\hat a\hat b\hat c} e^{\hat a}e^{\hat b}e^{\hat c}
\end{equation}
invariant under $G$. (This is only for illustration; the same holds with any number of legs.)
Applying the Leibniz rule to $\lie_\zeta T=0$ into the component expansion above,
one finds \begin{equation}
(\zeta_i)^\mu \partial_\mu (T_{\hat a\hat b \hat c})=0,
\end{equation} for $i=0,1,2$. This is equivalent to \eqref{derivatives-of-components}.
Combining \eqref{lie-formula} and \eqref{derivatives-of-components},
one finds that we have \begin{equation}
(\lie_{\xi_n} T)_{\hat a\hat b\hat c\cdots} =
(\xi_n)^{\hat i}_{,\hat a} T_{\hat i \hat b \hat c\cdots}+
(\xi_n)^{\hat i}_{,\hat b} T_{\hat a\hat i  \hat c\cdots}+\cdots
\end{equation} for our asymptotic Virasoro generators, i.e.~derivatives of 
components of $T$ do not appear. 

\subsection{Consequence of $t$--$\varphi$ reflection invariance}

One more trick uses the discrete symmetry of the background \eqref{metric}.
Note that it is invariant under the ``$t$--$\varphi$ reflection'' in the jargon of the black hole physics,
i.e. the transformation $t\to -t$, $\varphi\to-\varphi$. This inverts the time and the angular momentum
simultaneously, so it is not so unexpected that the black hole background is invariant under the reflection.
Now consider a two-index tensor $T_{ab}$ which is invariant under boost,
and even/odd under the $t$--$\varphi$ reflection. It is convenient again to introduce
$e^{\hat\pm} =e^{\hat t} \pm e^{\hat r}$. Then, of the components involving $\hat t$ or $\hat r$ directions,
the only invariant ones are $T_{\hat +\hat -}$ and $T_{\hat-\hat+}$ as argued in the
last section.
Moreover, the $t$--$\varphi$ reflection sends $e^{\hat\pm} \to -e^{\hat\mp}$.  One then has
\begin{equation}
T_{\hat +\hat -} = \pm T_{\hat - \hat +},
\end{equation} where $\pm$ depends on the even/odd-ness of $T$
under the $t$-$\varphi$ reflection. Note that this is a priori independent of the
(anti)symmetry under the interchange of two indices of $T$.
Converting to the indices $\hat r$ and $\hat t$, this means \begin{equation}
T_{\hat t\hat t}=-T_{\hat r\hat r}, \quad T_{\hat t\hat r}=0
\end{equation} for even $T$, and \begin{equation}
T_{\hat t\hat r}=-T_{\hat r\hat t},\quad T_{\hat t\hat t}=-T_{\hat r\hat r}=0
\end{equation} for odd $T$.

Let us apply this consideration to a four-index tensor $Z_{abcd}$
with the same symmetry as the Riemann tensor.
We consider the component $Z_{\hat t\hat r\hat\theta\hat\varphi}$ and related terms.
The first Bianchi identity implies \begin{equation}
Z_{\hat t\hat r\hat\theta\hat\varphi}+Z_{\hat r\hat\theta\hat t\hat\varphi}
+Z_{\hat\theta\hat t\hat r\hat\varphi}=0.
\end{equation}  Using the $t$--$\varphi$ reflection symmetry, one has \begin{equation}
Z_{\hat\theta\hat t\hat r\hat\varphi}=
-Z_{\hat t\hat\theta\hat r\hat\varphi}=Z_{\hat t\hat\varphi\hat r\hat \theta}.
\end{equation}
Therefore one obtains
\begin{equation}
Z_{\hat t\hat r\hat \theta\hat\varphi}=-2Z_{\hat t\hat\varphi\hat r\hat \theta}.
\end{equation}

\end{document}